\documentclass{aa}  

\usepackage{graphicx}
\usepackage{txfonts}
\usepackage{xcolor}
\usepackage{longtable}
\usepackage{lscape}
\usepackage{natbib,twoopt}
\begin{document}

   \title{Rapid Classification of TESS Planet Candidates with Convolutional Neural Networks}

   \author{H.P. Osborn\inst{1}
        \thanks{E-mail:hugh.osborn@lam.fr}\fnmsep\thanks{NASA FDL 2018 participant}
          \and M. Ansdell\inst{2}\fnmsep$^\dagger$
          \and Y. Ioannou\inst{3}\fnmsep$^\dagger$
          \and M. Sasdelli\inst{4}\fnmsep$^\dagger$
          \and\\
          D. Angerhausen\inst{5,6}\fnmsep\thanks{NASA FDL 2018 mentor}
          \and D. Caldwell\inst{7,8}\fnmsep$^\ddagger$
          \and J. M. Jenkins\inst{7}\fnmsep$^\ddagger$
          \and C. R\"{a}issi\inst{9}\fnmsep$^\ddagger$
          \and J. C. Smith\inst{7,8}\fnmsep$^\ddagger$
          }

   \institute{Aix Marseille Univ, CNRS, CNES, Laboratoire d'Astrophysique de Marseille, France
        \and Center for Integrative Planetary Science, University of California at Berkeley, USA
        \and Machine Intelligence Lab, Cambridge University, UK
        \and Australian Institute for Machine Learning, University of Adelaide, Australia
        \and Center for Space and Habitability, University of Bern, Switzerland
        \and Blue Marble Space Institute of Science, Seattle, USA
        \and NASA Ames Research Center, California, USA
        \and SETI Institute, California, USA
        \and Institut National de Recherche en Informatique et en Automatique, France
       }

   \date{Received February 22, 2019; accepted Month XX, 2019}

    \abstract
    {}
    {Accurately and rapidly classifying exoplanet candidates from transit surveys is a goal of growing importance as the data rates from space-based survey missions increases. This is especially true for NASA's \textit{TESS} mission which generates thousands of new candidates each month. Here we created the first deep learning model capable of classifying \textit{TESS} planet candidates.}
    {We adapted the neural network model of \citet{ansdell2018} to \textit{TESS} data. We then trained and tested this updated model on 4 sectors of high-fidelity, pixel-level simulations data created using the \textit{Lilith} simulator and processed using the full TESS SPOC pipeline. With the caveat that direct transfer of the model to real data will not perform as accurately, we also applied this model to four sectors of TESS candidates.}
    {We find our model performs very well on our simulated data, with $97\%$ average precision and $92\%$ accuracy on planets in the 2-class model. This accuracy is also boosted by another $\sim4\%$ if planets found at the wrong periods are included.
    We also performed 3- and 4-class classification of planets, blended \& target eclipsing binaries, and non-astrophysical false positives, which have slightly lower average precision and planet accuracies, but are useful for follow-up decisions.
    When applied to real TESS data, 61\% of TCEs coincident with currently published TOIs are recovered as planets, 4\% more are suggested to be EBs, and we propose a further 200 TCEs as planet candidates.}
    {}


   \titlerunning{TESS Planet Candidates from Neural Networks}
   \authorrunning{H.P. Osborn et al}
   \maketitle
%

\section{Introduction}
In the next two years NASA's Transiting Exoplanet Survey Satellite (\textit{TESS}) mission \citep{Ricker2014} is likely to more than double the number of currently known exoplanets \citep{sullivan2015transiting,huang2018expected,barclay2018revised}.
It will do this by observing 90\% of the sky for up to one year and monitoring millions of stars with precise enough photometry to detect the transits of extrasolar planets across their stars \citep[e.g.][]{huang2018tess,vanderspek2018tess,wang2018transiting}.
Every $\sim27.1$ day "sector" monitors the light of tens of thousands of stars which are then compiled into 1D "light curves", detrended for instrumental systematics, and searched for signals similar to transiting planets.
However, those signals with exoplanetary origin are dwarfed by signals from false positives --- those from artificial noise sources (e.g. systematics not removed by detrending), or from astrophysical false positives such as binary stars and variables.
Hence how best to classify exoplanetary signals is a key open question.

Answers until now include human vetting, both by teams of experts \citep{crossfield2018tess} or members of the public \citep{fischer2012planet}, vetting using classical tree diagrams of specific diagnostics \citep{mullally2016identifying}, ensemble learning methods such as random forests \citep{mccauliff2015automatic,armstrong2018automatic}, and deep learning techniques such as neural networks \citep{Shallue2018,schanche2018machine,ansdell2018}.
The current process of vetting \textit{TESS} candidates involves a high degree of human input. In \citet{crossfield2018tess}, 19 vetters completed the initial vetting stage of 1000 candidates (henceforth TCEs, or Threshold Crossing Events), with each candidate viewed by at least two vetters.
However each \textit{TESS} campaign has so far produced more than 1000 TCEs, and a simple extrapolation suggests as many as 500 human work hours may be required per month to select the best \textit{TESS} candidates.


The first attempts at classification using neural networks have tended to use exclusively the lightcurve \citep[e.g.][]{Shallue2018,Zucker2018}.
In \citet{ansdell2018}, we modified the 1D light-curve-only neural network approach to candidate classification of \citet{Shallue2018} to include both centroids and stellar parameters, subsequently improving the precision of classification.
In this paper we show the results of adapting those models to both simulated and real \textit{TESS} data, the first time deep learning has been performed for \textit{TESS} data.



\section{Datasets} 
\label{sec:data}

\subsection{TSOP-301}
\label{sec:simulated-data}
As no flight data existed at the start of the project, we relied on multiple end-to-end simulations performed by the \textit{TESS} team. Three such runs were considered for use: an initial 1-sector run named ETE-6 \citep{jenkins2018simulated} used for the final \textit{TESS} mission ground segment integration test, a 2.5-sector run (2 whole sectors and a further sector including only the overlap region) named TSOP-280 used for the final validation and verification of the TESS science processing pipeline \citep{2016SPIE.9913E..3EJ} and a 4-sector run called TSOP-301\footnote{Called ``TSOP-301'' from the \textit{TESS} operations issue tracking ticket which initiated the run.}  which was specifically designed to create a test set for machine learning and to characterize detection characteristics of the \textit{TESS} pipeline. We focused on the TSOP-301 run, which had the most data and the most complete set of simulated features.

TSOP-301 was a full 4-sector end-to-end run of the \textit{TESS} science processing pipeline. To help facilitate the development of the Science Processing Operations Center (SPOC) pipeline, it was necessary to produce simulated flight data with sufficient fidelity and volume to exercise all the capabilities of the pipeline in an integrated way. Using a physics-based \textit{TESS} instrument and sky model, the simulation tool, named \textit{Lilith} (Tenenbaum et al. 2019, in preparation), creates a set of raw TESS data which includes models for the CCDs, readout electronics, camera optics, behavior of the attitude control system (ACS), spacecraft orbit, spacecraft jitter and the sky, including zodiacal light, and the TESS Input Catalog (TIC). The end product being an array of expected instrumental artifacts and systematics (e.g. cosmic rays, CCD readout errors, thermal-induced focus errors and spacecraft jitter-induced pointing errors). The model also incorporates realistic instances of stellar astrophysics, including stellar variability, eclipsing binaries, background eclipsing binaries, transiting planets and diffuse light. 

This simulated raw image data set is then passed through the SPOC pipeline providing full integration tests of the science processing from raw pixel calibration, to transiting planet search \citep{2010SPIE.7740E..0DJ,2013ApJS..206...25S}, to the generation of archivable data products such as tables of TCEs and Data Validation products \citep{2018PASP..130f4502T,2019PASP..131b4506L}. Full instrumental and astrophysical ground truth is generated for each \textit{Lilith} run and can be used as a training set. 

\begin{figure}
\begin{centering}
\includegraphics[width=\columnwidth]{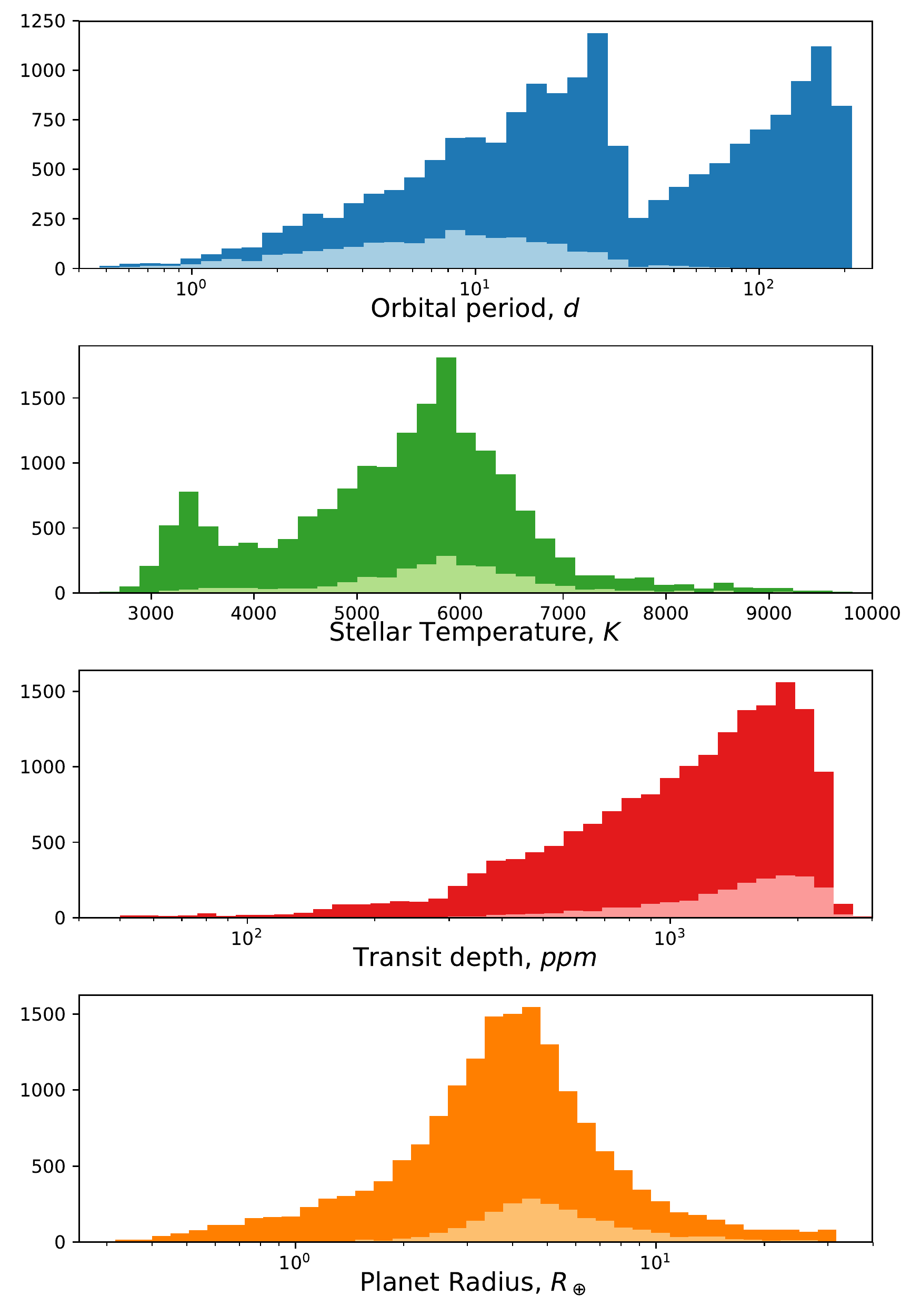}
\caption{Distribution of injected planet signals as a function of key inputs. The split distribution in the upper plot is due to an injected distribution into the multi-sector regions that was flat in linear period space. The lighter colour shows those injections present in TCEs.}
\label{fig:injdists}
\end{centering}
\end{figure}

In TSOP-301 we simulated 4-sectors using the then current \textit{TESS} target management and selection with use of the TESS Input Catalog version 6. There were 16,000 targets per sector and many targets were near the ecliptic pole resulting in many targets observed for more than one sector.
Realistic planet distributions based on current understood planet populations was not used, instead a distribution was generated with good overlap with the desired \textit{TESS} planet detectability in order to give a machine learning classifier a good distribution of signals to train on. 
20\% of all targets had planetary transits, the distributions for which are seen in Figure \ref{fig:injdists}. 
An additional 20\% had eclipsing binaries (EBs) or background eclipsing binaries (BEBs) in order to give the classifier a good set of potential astrophysical false-positives.
Using appropriate dataset balancing (i.e. Section \ref{sec:balancing}), this difference should have minimal effects on the performance of a deep learning model.

\subsection{Pre-Processing} 
\label{sec:processing}
\textit{TESS} lightcurves were pre-processed in a method similar to that of \citet{Shallue2018} and \citet{ansdell2018}, iterating over each TCE to produce binned phase-folded "global" (full lightcurve showing the entire phase between $-0.5$ and $0.5$) and "local" (zoomed in on the transit between $-2.5t_{\rm dur}$ to $2.5t_{\rm dur}$) views of both the light and centroid curves (see Figure 3 in \citealt{Shallue2018} or Figure 1 in \citealt{ansdell2018}).
A lightcurve with instrumental systematics removed is compiled for each target before it is searched for transiting planets (the so-called Pre-search Data Conditioning, or PDC, light curve), and then a lightcurve with all non-planetary signals detrended is produced after a candidate detection has been found (the Data Validation, or DV, light curve).
Unlike \citet{Shallue2018} and \citet{ansdell2018}, which used exclusively the PDC lightcurves, we used both of these time series. These were accessed both from the \textit{TESS} MAST pages\footnote{\url{https://archive.stsci.edu/tess/bulk_downloads.html}}
 
The DV time series contain unique time series for each candidate planet, with flux during the transits of previously detected TCEs removed.
We took the detrended (\texttt{LC\_DETREND}) DV lightcurve where these were available, using the initial (\texttt{LC\_INIT}) lightcurve if not.
Centroid information is found exclusively in the PDC files, and comes in two types --- PSF centroids (which is calculated using a model of \textit{TESS}'s point spread function) and the MOM centroids (which is simply the weighted centre of light within the \textit{TESS} aperture).
We extracted row and column PSF centroids where available, as these are typically more accurate, but reverted to MOM when these were missing. In both cases, the median was subtracted giving relative x- and y- shifts in the centre of light.

Anomalies greater than 3.5-sigma from surrounding points are also removed from each time-series.
Then both time series were phase folded and median binned into global and local views.
We primarily use the DV lightcurve for the final "views" of the TCEs, however in some cases, the gaps around previously detected transits cause large gaps in the final views. We pick a threshold of 50\%, above which the PDC lightcurve views are instead used.
These lightcurves are then normalised using the detected depth such that the median out-of-transit is 0.0 while the transit depth is at -1.0.
The row and column centroids are first added in quadrature, and then also phase folded and binned into global and local views.
To normalise the centroids, the out-of-transit median is subtracted, the time series is then inverted (to match the "flux drop" of a transit), and finally it is multiplied by the ratio between the out-of-transit RMS in the (normalised) light curve and this new centroid curve. 
This is done to make centroid curves with no significant deviation a flat line (rather than amplify low-significance structure).

Some TCEs remain with large numbers of gaps in the phase-folded views (due to detection near or on gaps in the lightcurve), which is problematic as "NaNs" are undifferentiable, and therefore cause immediate errors when present in data seen by a machine learning model.
Models with NaNs and without anomalies removed initially struggled to train, likely as a result of too many objects with missing data.
To avoid this, we remove 4577 TCEs for which more than 25\% of the local view is missing.
Only 4\% of these constituted planetary candidates, therefore the overall fraction of planets actually increased due to this cut.
We also filled any missing data in the remaining views with zeros (which matches the median out-of-transit value), although our white noise augmentation step (see section \ref{sec:augmentation}) means the model sees Gaussian noise for these missing values. 

\subsection{Stellar and Transit Parameters}
\label{sec:starpars}
The neural networks will classify data using the shape and distribution of the input transit data.
However, extra information can be found by using other parameters which may also help classification.
This includes stellar parameters, which testing in \citet{ansdell2018} showed provided a boost of around 0.5\% in accuracy for planet classification (potentially as a result of identifying large stars unlikely to be planet hosts).
However, the planetary injections performed by \textit{Lilith} effectively choose random stars rather than following any physical correlations (such as trends in planet occurrence with metallicity or stellar mass), therefore stellar parameters are unlikely to provide as big a boost.
Some transit phenomena may also not be represented in the lightcurve data but may aid classification, the most obvious being depth and duration --- both an overly deep and an overly long eclipse may suggest an eclipse of two similar-sized objects. However one or both of these are lost during global and local view normalisation.
We therefore added the following additional data:
From the transit search: the orbital period, transit duration, the semi major axis scaled to stellar radius ($a/R_s$), the number of transits $N_{\rm TRANS}$, the transit SNR, the transit depth and ingress duration. 
Derived from the transit model fit parameters we added the radius ratio $R_p/R_s$, the impact parameter $b$, the ratio of the maximum MES \citep[multiple event statistic, a proxy for SNR,][]{Jenkins2002} to the expected MES from the SES (single event statistic; i.e. the SNR of the strongest signal) ${\rm SES}\sqrt{N_{\rm trans}}$, the logged planet radius divided by a arbitrary planetary boundary (set at 13$R_\oplus$), and the logged ratio of the transit duration over the expected duration for a planetary transit given the stellar density and orbital period.
And from stellar parameters we added the \textit{TESS} band magnitude, stellar radius, total proper motion, stellar $\log{\rm g}$, stellar metallicity, and stellar effective temperature.
We took these values from the DV lightcurve fits headers provided for each TCE. 

All this additional data was then normalised by subtracting the median and dividing by the standard deviation.

\subsection{Labels}
Unlike for real flight data, the ground truth of our simulated \textit{TESS} dataset is known precisely. However, the injected signals are never recovered perfectly during transit search --– some may be found at the wrong period, or with the wrong durations, etc.
Therefore, the degree of correlation between the injected signal and the recovered TCE must be computed --- we adapted the \textit{TESS} team’s code which sums in quadrature the number of cadences which overlap between the in-transit (or in-eclipse) points from an injection and from the detection, setting a threshold of 0.75.

We split eclipsing binaries into their primary and secondary dips, therefore recovering both signals.
We also searched for injections recovered at an integer multiple of the real period, finding a handful of equal-depth eclipsing binaries detected at half the real period.
Although complex labels were generated for each target (for example, EB\_secondary or BEB\_at\_2P), we collated all labels from the same source to give between four (Planet, Eclipsing Binary, Background Eclipsing Binary, Non-Astrophysical signal) and two ("planet" \& "not planet") labels, depending on the model used.

\subsection{TESS sectors 1 to 4}
\label{sec:real-data}
At the point of submission, data from four \textit{TESS} sectors have been released. 
TCE catalogues have been compiled from the $\sim16000$ 2-minute cadence targets observed each sector\footnote{\url{http://archive.stsci.edu/tess/bulk_downloads/bulk_downloads_tce.html} for sectors 1-3, and we took the information in the released DV lightcurves to build a TCE catalogue for campaign 4}. In total, this gives 7562 TCEs from 3266 unique \textit{TESS} IDs, which includes duplications between sectors.
Of these, 370 have been published as TOIs (\textit{TESS} Objects of Interests). Their identification comes from candidates identified by the "quick look pipeline" \citep{2018AAS...23143909F} which are then manually vetted in the manner of \citet{crossfield2018tess}.

The \textit{TESS} lightcurves were processed in the same way as for the simulated data (see Section \ref{sec:processing}). We also performed the same removal of lightcurves which had more than 20\% of points in either of the phase-folded views missing. This led to the removal of 2197 candidates.

Despite being generated from the pixels with realistic noise sources, the simulated data is unlikely to be identical to the real data in key ways, especially in terms of unexpected systematic noise sources. This likely includes the second orbit of sector 1 which has higher than average systematic noise due to unexpected noise in fine pointing. However, some injected noise sources have been identified as not present in the real data, such as the sudden pixel sensitivity dropouts (SPSDs) which were present in \textit{Kepler}.



\section{Machine Learning Models} 
\label{sec:models}


\begin{figure}
\begin{centering}
\includegraphics[width=\columnwidth]{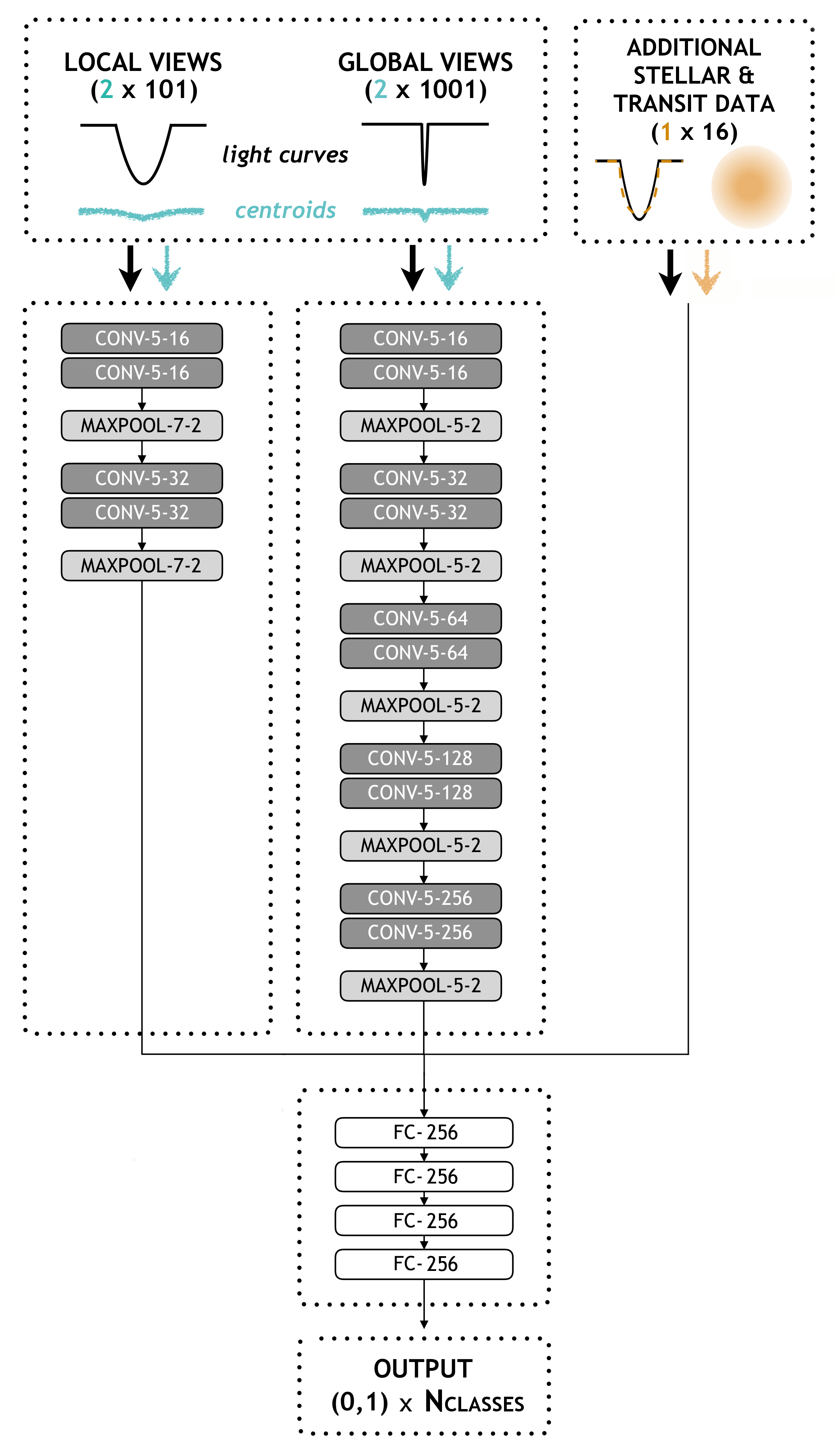}
\caption{The convolutional neural network architectures used in this work. "CONV": A 1D convolutional layer, with the two following numbers referring to the kernel size and the number of filters; "MAXPOOL" refers to the process of 1D max pooling the tensor, and the numbers refer to kernel size and stride; "FC" is a fully connected, or linear, ANN layer where the number shows the number of neurons.}
\label{fig:architecture}
\end{centering}
\end{figure}

\subsection{Architecture}
In \texttt{Astronet} \citep{Shallue2018} and \texttt{exonet} \citep{ansdell2018}, a series of convolutional layers are applied to the local and global views, with the larger global view having a deeper convolutional structure.
These are then combined together as inputs for a series of fully connected layers before outputing a single class prediction. Figure \ref{fig:architecture} gives an overview of the model architecture.

We maintained the convolutional filter sizes and architecture from \texttt{Astronet}, with four 1-dimensional convolutional layers for the local view, and 8 for the global view.
Every two layers, max pooling is performed to reduce the overall size of the tensor.
With the number of input data points shrunk by a factor of 2 (see section \ref{sec:inputdims}), the final fully connected layers were similarly shrunk from 512 to 256 neurons.
The dimensionality of the output depends on the model loss function, with either a single prediction per object (binary) or a prediction per class, per object (multi-class)

For binary models, the binary cross entropy loss function (\texttt{BCEloss} in \texttt{pytorch}) was used, whereas for multi-class models, a Cross Entropy loss (\texttt{CrossEntropyLoss} in \texttt{pytorch}) function was used.
For gradient descent, we used Adam \citep{kingma2014adam} as an optimizer with a starting learning rate around $2\times10^{-5}$.

In all cases, we trained until the output of the loss function when applied to validation data had stopped decreasing; a sign that the model is well-fitted but not yet begun to over-fit. This was between 200 and 500 epochs, depending on the learning rate and number of classes used.

\subsection{Balanced Batch Sampling}
\label{sec:balancing}
Training a neural network using a dataset with an unbalanced class distribution is difficult \citep[see, e.g.][]{chawla2004special}, since the learning algorithm inevitably biases the model towards learning the majority class.
In the case of the \textit{Kepler} dataset used in \citet{ansdell2018}, the two classes (planet and non-planet) were more closely balanced than the data here. This was partly because candidates labelled as "unknown" by human vetters \citep{batalha2013planetary,2014ApJS..210...19B,2015ApJS..217...31M,2015ApJS..217...16R} were classified and removed from the DR24 sample.
However, such a step is not available with our \textit{TESS} dataset, hence only 14\% of the TCE dataset are planets. It is therefore necessary to perform dataset balancing in order to train the network. 
We took an approach that involves resampling the input data (rather than, for example, weighting the loss function).
We did this by balancing the mini-batches used in training, meaning each training epoch sees an equal number of samples from each class \citep[see, e.g.,][]{he2008learning}.

\subsection{Cross Validation}
To test the model while retaining as much of the data as possible for training, we used cross validation.
This splits the data into $k$ parts, and independently trains such that a different subsection of data is kept as validation data each time, while $(k-1)$ parts are used for training.
We used $k=8$ for all models here, to utilise all available GPUs\footnote{As provided by our google cloud education grant, \url{https://cloud.google.com/edu/}}.

\subsection{Augmentation}
\label{sec:augmentation}
Augmentation is the process of modifying training data in order to generate similar but not identical samples, thereby increasing the effective number of samples.
This therefore helps preserve against over-fitting.
We used three methods of augmentation: White Gaussian noise was added to each light and centroid curve, with the amount chosen randomly between 0 and the out-of-transit RMS of each light and centroid-curve; A shift of between -5 and 5 bins was applied to the phase; and  50\% of all time series were mirrored in phase.
These were tested using cross validation on a baseline binary model to assess whether augmentation does improve model training, with the results seen in Table \ref{tab:testing}.
It was found that adding each improved the overall precision of the model, with the removal of Gaussian white noise augmentation having the largest effect (~3.1\% decrease in A.P.).

\subsection{Input array dimensions}
\label{sec:inputdims}
For both the \texttt{astronet} and \texttt{exonet} models, input arrays of size 2001 and 201 were used for global and local views respectively.
Reasons for this included that long period planet candidates seen with \textit{Kepler} needed at least a single bin on the global view, and that high resolution local view allows the in/egress of small planets to be resolved.
\textit{TESS}, which will find shorter period planets which are on average larger than those of \textit{Kepler}, therefore may not need such wide bins.
We tested whether reducing the number of bins by a factor of 2 improved performance with \textit{TESS} (see Table \ref{tab:testing}). 
This shows that a smaller lightcurve view does indeed improve model performance, likely because increasing SNR in each bin outweighs the effects from low phase resolution in \textit{TESS}.
Halving the number of bins also increases run speed.

\begin{table}
    \caption{Results on testing different model augmentation and input view sizes. The 101 and 201-bin models are with all three methods of data augmentation. Testing of individual augmentation techniques was performed by removing each individual method in turn from the 101-bin model. 4-fold cross validation was used for this testing, and the numbers given are on the validation dataset.}
    \label{tab:testing}      
    \centering                          
    \begin{tabular}{|l|c|c|}        
        \hline                 
        \textbf{Model} & \textbf{Avg. Precision} \\    
        \hline                        
        201/2001 bins & $92.0\pm0.7\%$  \\
        101/1001 bins & $92.7\pm0.7\%$  \\
        \hline
        Without white noise & $89.6\pm0.7\%$  \\
        Without phase-inversion & $90.4\pm0.7\%$  \\
        Without phase-shifts & $90.5\pm0.7\%$  \\
        Without any augmentation & $85.2\pm0.7\%$  \\
        \hline                        
    \end{tabular}
\end{table}


\section{Results With Simulated Data}
\label{sec:simulation_res}

To best assess the accuracy and precision of each model, we performed an "ensemble", or bagging, method by taking all 8 models trained during cross validation and applying these to the test data taking the mean class prediction across all.
Ensembles typically outperform single models and guard against models which may find local minima \citep[see, e.g.][]{dietterich2000ensemble}.
Although usually different initialisation weights are used for each ensembled model, we used the same random initialisation weights. However, a test with the binary model confirms that, due to each model seeing different training and validation datasets, there is no difference in performance.
We used the 10\% of data which was randomly left out of the train/validation set.
These results are shown, for each model, in Table \ref{tab:modelresult}.

Here we define accuracy as the fraction of all predicted class members that are correct (${\rm TP}/({\rm TP}+{\rm FP})$, often also called precision); recall as the fraction of all objects of a class which are correctly predicted (${\rm TP}/({\rm TP}+{\rm FN})$); and average precision as the average accuracy (or precision) for all classes, weighted by the class frequency (a so-called micro average, implemented with \texttt{scipy}'s \texttt{average\_precision\_score})\footnote{The micro-averaged average precision simplifies to $({\rm TP}+{\rm TN})/({\rm TP}+{\rm TN}+{\rm FP}+{\rm FN})$ in the binary case}.

We find the binary model gives the best planet accuracy, while the three class model gives both the highest average precision on planets and on all classes.

\begin{table}
    \caption{Accuracy, recall and average precision for the trained model on the test set, using a mean of the predictions from all 8 ensemble models. For overall average precision, a "micro" average is performed which better account for imbalanced datasets.}
    \label{tab:modelresult}
    \centering
    \begin{tabular}{|l|c|c|c|}
        \hline
         & Accuracy & Recall & Average Precision \\
        \hline
        \multicolumn{3}{|c|}{Binary} & 97.3\\
        \hline
        Planet & 91.8 & 87.8 & 95.2 \\
        Not Planet & 97.6 & 98.5 & 99.4 \\
        \hline
        \multicolumn{3}{|c|}{3-class} & 97.1 \\
        \hline
        Planets & 90.4 & 90.1 & 95.6 \\
        EBs & 95.1 & 95.1 & 96.9 \\
        Unknown & 94.8 & 94.9 & 97.7 \\
        \hline
        \multicolumn{3}{|c|}{4-class} & 96.3\\
        \hline
        Planets & 89.1 & 88.8 & 94.4 \\
        EBs & 87.4 & 91.7 & 94.7 \\
        BEBs & 88.5 & 81.7 & 91.7 \\
        Unknown & 94.6 & 95.5 & 97.8 \\
        \hline
    \end{tabular}
\end{table}

\begin{figure}
\begin{centering}
\includegraphics[width=\columnwidth]{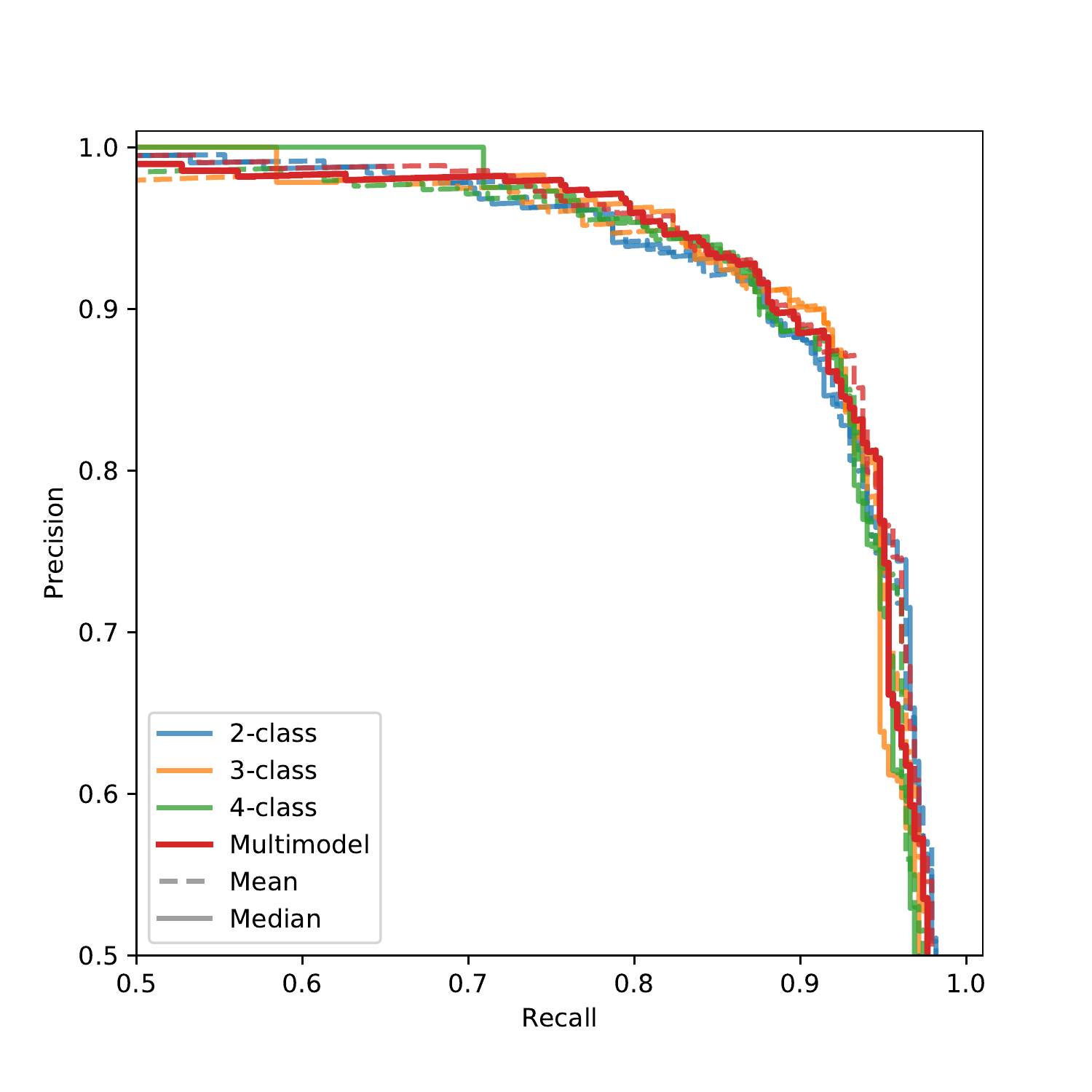}
\caption{Precision-recall curve for planets in all three models. Random guessing would produce a straight line at the planet frequency ($\sim 14\%$), which is far below the base of this zoomed in plot. Perfect models are as close as possible to the top right corner.  We show both mean (solid lines) and median (dashed lines) for the three models (2,3,4 refer to binary, 3-class and 4-class). We also plot "Multimodel" averages which ensemble the planet prediction from all three of those models. }
\label{fig:compPR}
\end{centering}
\end{figure}

\begin{figure}
\begin{centering}
\includegraphics[width=\columnwidth]{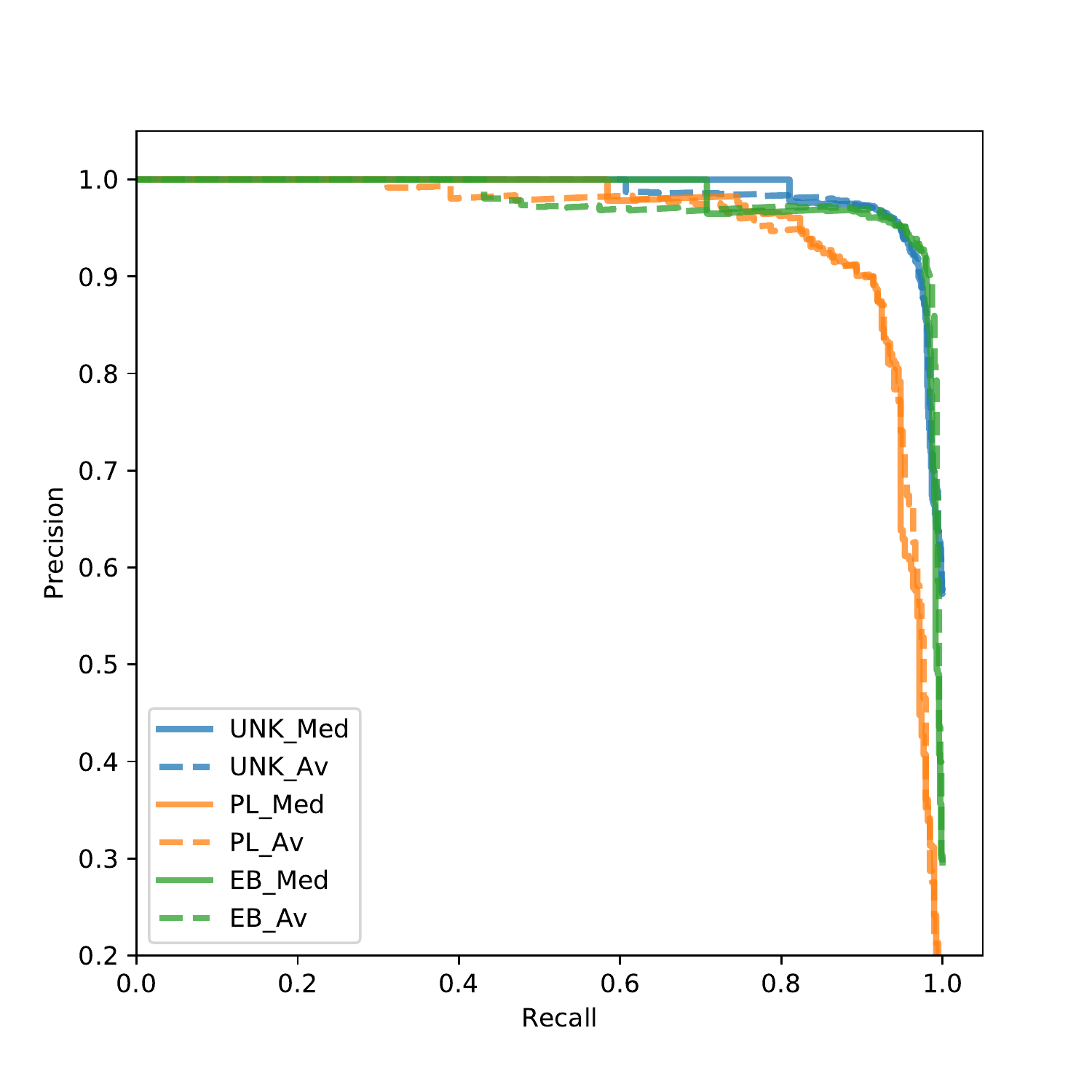}
\caption{Precision-recall curve for our 3-class Model, with both median (Med) and mean (Av) averages. "UNK" refers to unknown, or non-astrophysical sources; PL refers to planets; and EB refers to Eclipsing Binaries.}
\label{fig:3classPR}
\end{centering}
\end{figure}

\begin{figure}
\begin{centering}
\includegraphics[width=\columnwidth]{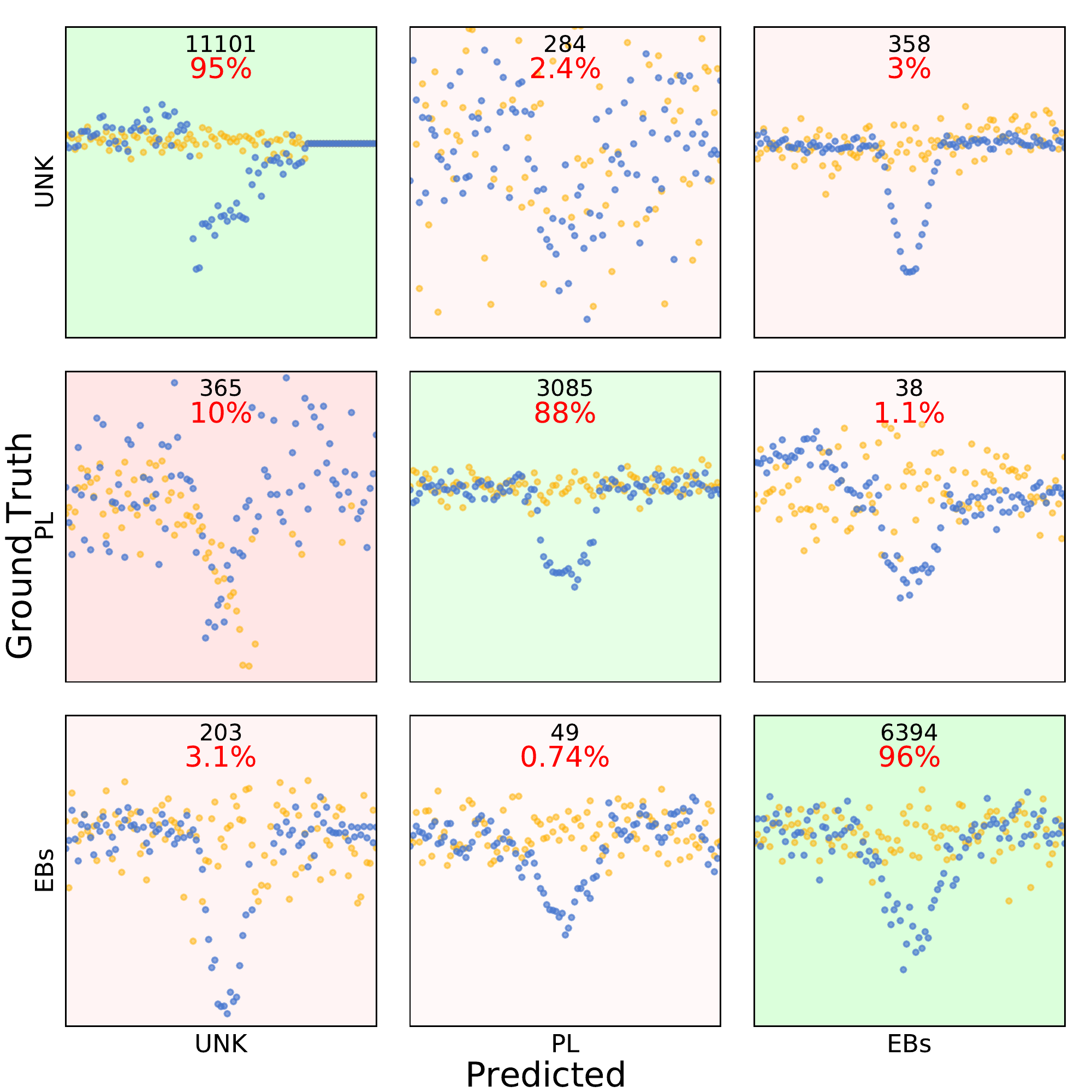}
\caption{Confusion Matrix, using real data from the cross validation dataset, for the 3 class model. Those classes are "UNK" from non-astrophysical or unknown sources, "PL" from planetary transits and "EB" from eclipsing binaries. The binned \& phase-folded input data from a single randomly-selected object is shown in each subplot, with the light curve in blue and the centroid curve in orange. The black number is the total number of objects classified in this subset, while the red number shows recall to two significant figures, i.e. what proportion of each class is prediction to be in this class (hence horizontal rows always sum to 100\%, within rounding errors). Objects on the diagonal are correctly predicted and coloured green, while those outside are mis-classified and coloured red. The strength of those background colors is proportional to the percentage (i.e. recall) in each box. }
\label{fig:3classCM}
\end{centering}
\end{figure}

\begin{figure}
\begin{centering}
\includegraphics[width=\columnwidth]{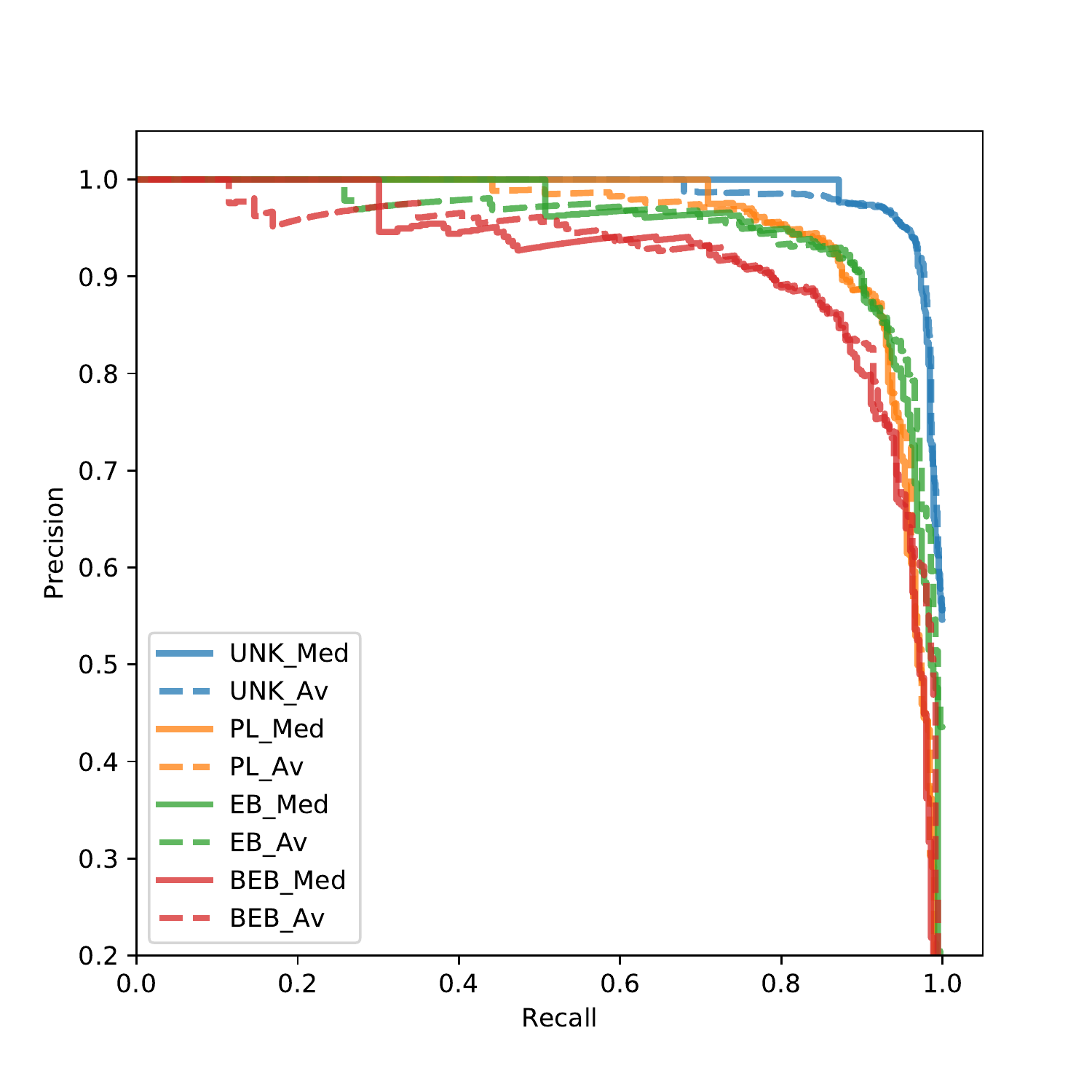}
\caption{Precision-recall curve for the 4 class \texttt{exonet} model. The EB and BEB models perform poorly in comparison to Figure \ref{fig:3classPR}, mostly due to confusion with each other (see Figure \ref{fig:4classCM}).}
\label{fig:4classPR}
\end{centering}
\end{figure}

\begin{figure}
\begin{centering}
\includegraphics[width=\columnwidth]{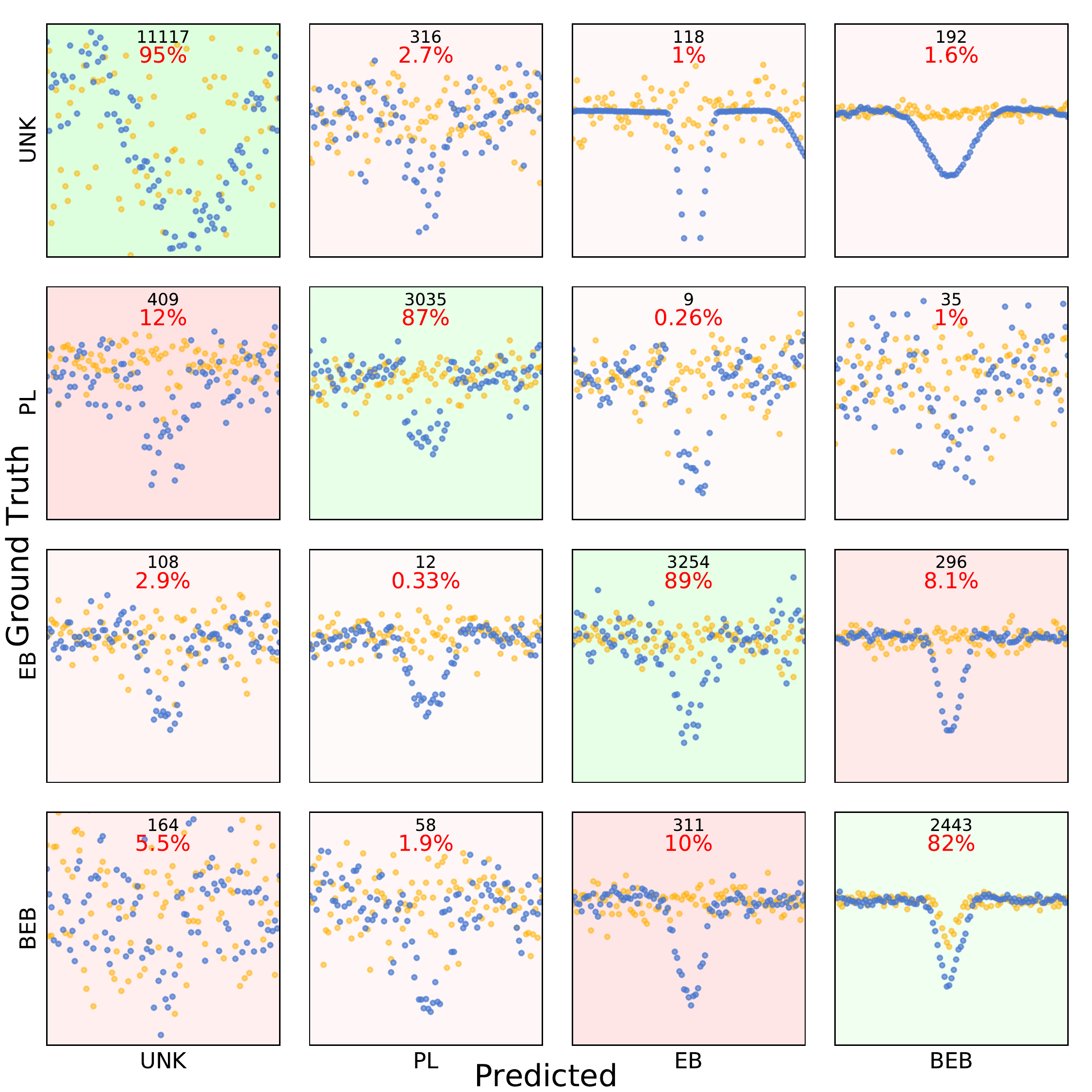}
\caption{Confusion Matrix, as Figure \ref{fig:3classCM}, for the 4-class \texttt{exonet} model.}
\label{fig:4classCM}
\end{centering}
\end{figure}

\begin{figure}
\begin{centering}
\includegraphics[width=\columnwidth]{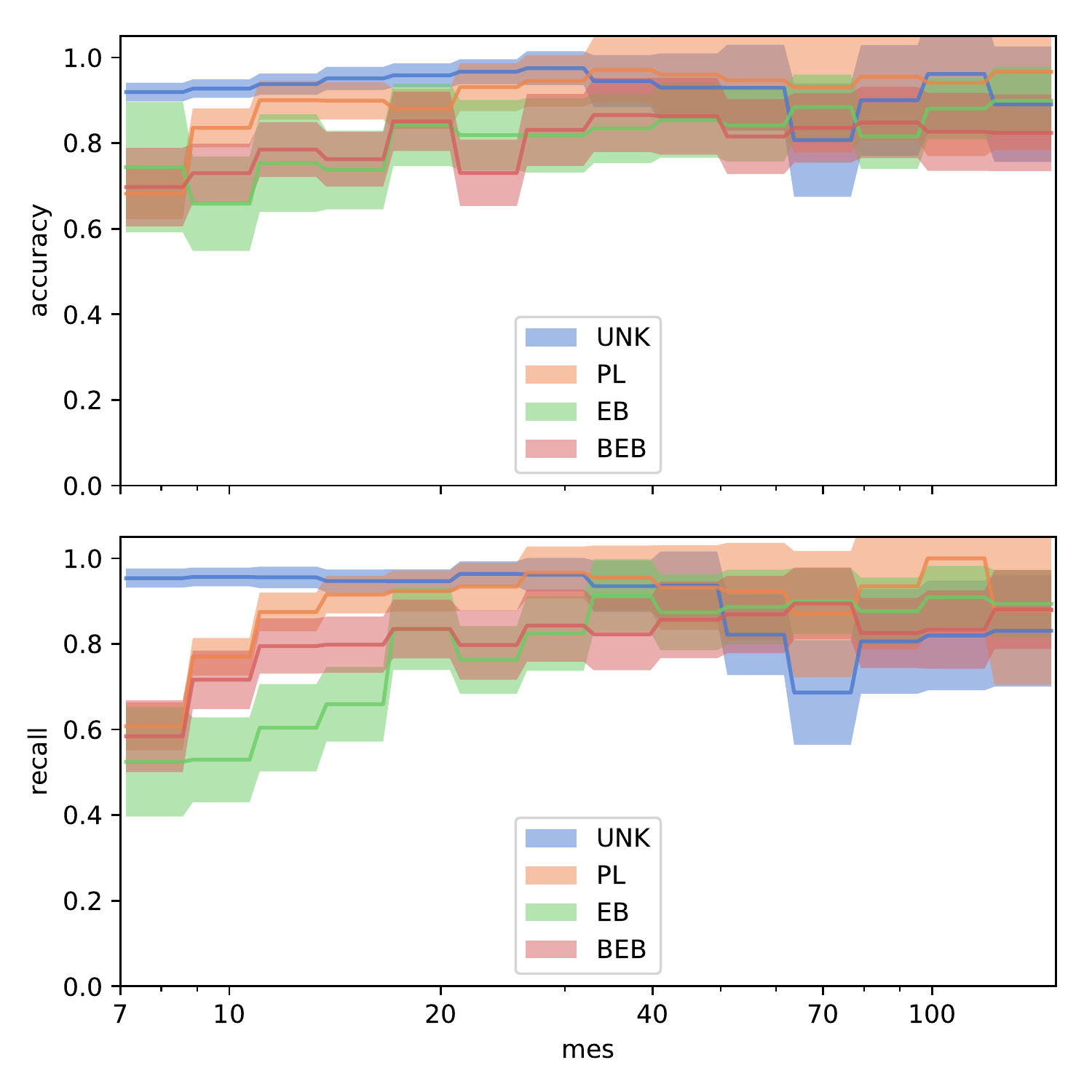}
\caption{Comparison of recall and accuracy as a function of the Multiple Event Statistic (effective SNR from the transit planet search) for each of the classes in our 4-class model.}
\label{fig:MESplot}
\end{centering}
\end{figure}

In Figure \ref{fig:compPR} we show a comparison of the precision-recall curve for all three class categories on exclusively planets. These show near perfect agreement suggesting the addition of other classes do not inhibit a model's ability to differentiate planets.
In Figures \ref{fig:3classPR} and \ref{fig:4classPR} we show the PR curves for each class in the multi-class 3- and 4-class models respectively, when applied to the test dataset. We calculate both the median and mean values across all eight ensemble models. Due to the tendancy of predictions to cluster near 0 or 1, the median gives higher precision better at more restrictive thresholds (e.g. low recall) while the mean gives higher recall for less restrictive thresholds (e.g. low precision)\footnote{This is most likely due to the sigmoid layer, which distributes class predictions close to either 0 or 1. For example, the mean of predictions [0,0,0,0,0,1,1,1] is more inclusive for misidentified candidates (0.375) that a median (0.0).}.
In Figure \ref{fig:3classCM} and \ref{fig:4classCM} we show confusion matrices for the 3- and 4-class model using cross validation data, including randomly selected local view light curves for each class.
In Figure \ref{fig:MESplot} we compare the performance of recall and accuracy with respect to MES (Multiple Event Statistic, a proxy for the signal SNR) using cross validation data for the 4-class model.



\section{Application to Real TESS Data}
\label{sec:real_res}

We directly applied the trained models to those TCEs from the first four sectors of real \textit{TESS} data (see Table A.\ref{tab:toipreds}).
As well as using the ensembled models from section \ref{sec:simulation_res}, we also compiled an average planet class probability by averaging all three models.

To check how well the model was performing, we loaded the TOIs so far published on the TESS alerts database\footnote{\url{https://tess.mit.edu/alerts/}, accessed 2019/02/08}, which is compiled using candidates from both the SPOC pipeline (which come from the TCE tables used here) and the so-called quick-look pipeline (QLP).
Planet candidates are then identified by manual vetting as in \citet{crossfield2018tess}.
As expected, our TCE list contains all but one of the 146 SPOC-derived TOIs, but only 46 of the 207 QLP-derived TOIs.
Including duplications due to candidates identified in multiple sectors, we found 353 TCEs which corroborated with 201 TOIs (based on a combination of period, epoch, duration and depth matches).
61\% of those TCEs are predicted as planets with a threshold of >50\% in our average classifier. 
Ignoring duplications and taking the classification from the most sectors (or otherwise averaging the classifications) gives 112 out of 212 TOIs in agreement.

Of the 43 planets known before launch (the majority being hot jupiters), 95.2\% were classified as planets by the model, with the exception of WASP-18b ($p_{\rm pl}=49.9\%$) which has a detectable secondary eclipse \citep{shporer2018tess}, and HATS-34b ($p_{\rm pl}=39\%$) which has a V-shaped transit \citep[$b=0.94$][]{de2016hats}.

Of the 14 planets already confirmed by \textit{TESS} from the TOI list, we identify TOIs 123b, 135b, 125b, 120b, 125c and 174b strongly ($p_{\rm pl}>0.9$), weakly recognise TOIs 216b, 197b, 144b and TOI-136b ($0.3<p_{\rm pl}<0.9$) and misidentify TOI-125c, TOI-216c, and TOI-256b \& c.

15 more are ranked as EBs or BEBs (with $p>0.5$ in either model), which we list in Table \ref{Table:EBs}.
However, a quick manual vetting of these signals does not come to the same conclusion, with TIDs 2760710, 92226327, 231702397 and 307210830 still possible planet candidates.
95 and 82 objects are classed as "unknown" (e.g. from a non-astrophysical source) with the 3- and 4-class models respectively.

\begin{table}
\caption{TOI objects with a high likelihood of being astrophysical false positives. We only show all classes for the 3-class model, and the split EB and BEB classes from the 4-class model. $\dagger$ marks HATS-34b.}
\label{Table:EBs}
{\tiny
\begin{tabular}{|l|l|r|r|r|r|r|}
\hline
       TID &     toi & UNK$_3$ &  PL$_3$ &  EB$_3$ & EB$_4$ & BEB$_4$ \\
\hline
   2760710 &  227.01 &          0.090 &         0.000 &         0.910 &         0.836 &          0.024 \\
 279740441 &  273.01 &          0.574 &         0.000 &         0.426 &         0.000 &          0.835 \\
 425934411 &  142.01 &          0.506 &         0.001 &         0.493 &         0.000 &          0.678 \\
 272086159 &  176.01 &          0.001 &         0.000 &         0.999 &         0.879 &          0.003 \\
 272086159 &  176.01 &          0.002 &         0.000 &         0.998 &         0.800 &          0.000 \\
 237924601 &  252.01 &          0.006 &         0.000 &         0.994 &         0.003 &          0.873 \\
  92226327 &  256.01 &          0.150 &         0.000 &         0.850 &         0.558 &          0.276 \\
 425934411 &  142.01 &          0.157 &         0.005 &         0.838 &         0.000 &          0.979 \\
 425934411 &  142.01 &          0.033 &         0.000 &         0.966 &         0.000 &          0.853 \\
 237924601 &  252.01 &          0.021 &         0.107 &         0.872 &         0.125 &          0.592 \\
 231702397 &  122.01 &          0.199 &         0.390 &         0.411 &         0.057 &          0.640 \\
 307210830 &  175.02 &          0.186 &         0.412 &         0.401 &         0.000 &          0.610 \\
 176778112 &  408.01 &          0.352 &         0.014 &         0.634 &         0.000 &          0.520 \\
 355703913 &  111.01$^\dagger$& 0.055 &         0.010 &         0.935 &         0.005 &          0.408 \\
 237924601 &  252.01 &          0.110 &         0.399 &         0.491 &         0.011 &          0.515 \\
\hline
\end{tabular}
}
\end{table}

Interestingly, a further 200 TCEs not classified as TOIs are predicted as members of the planet class (see Table A.\ref{tab:realpreds}).
These are spread across 144 unique \textit{TESS} objects, with 57 of those having a class probability greater than 90\%.
After viewing these 200 TCEs, we plotted a handful of the most promising planetary candidates in Figure \ref{fig:NewPLs}. 




\section{Discussion}

\subsection{Comparison with Ansdell, 2018}
In \citep{ansdell2018}, we achieved an average precision of 98\%, with an accuracy on planets of 97.5\%.
In this study, we are unable to achieve such a high average precision or accuracy, with 97.3\% average precision for the binary model and 92\% accuracy on planets in the 3-class model.
A number of limitations could explain this discrepancy, which we cover in turn here.

The most obvious is in the presence of more false positives in the \textit{TESS} input data, whereas some non astrophysical false positives (objects labelled as unknown by \citet{batalha2013planetary,2014ApJS..210...19B,2015ApJS..217...31M,2015ApJS..217...16R} were removed from the samples in both \citet{Shallue2018} and \citet{ansdell2018}.
The abundance of non-astrophysical false positives in this TESS dataset may also be caused in part by the reduction in minimum number of transits from three to two, allowing two non-periodic noise sources to combine to give a candidate signal (far more difficult in the $N_{\rm t}>\geq3$ case).
Another discrepancy is in the source of labels: for this study the ground truth was known absolutely, for the most part, thanks to simulations. 
In the \textit{Kepler} dataset, only those signals identified as planet candidates by humans were positive classes, introducing a possible human bias.
For example, more difficult to identify planets (e.g. those affected by systematic noise) may have been missed in human vetting, improving the overall quality of the planet class.

It may seem like another difference might be the proportion of low-SNR planet candidates in \textit{TESS} compared to \textit{Kepler}, which may be intrinsically higher due to the larger average flux uncertainties.
However, this is not the case, and the distribution of injected \textit{TESS} planets \& \textit{Kepler} planet candidates is similar in terms of SNR.
Instead the goalposts have been shifted and low-SNR \textit{TESS} candidates are found, on average, at a larger planetary radius. 
This itself may be problematic, as large planets are more easily confused with eclipsing binaries, although the difference is likely minimal.

Another significant difference between the \textit{TESS} and \textit{Kepler} datasets is in the centroids.
The uncertainty in the centre of light (i.e. centroid) is determined by two things - the total number of photons and the number of pixels that light is spread over.
\textit{TESS} suffers in both of these cases when compared to \textit{Kepler}, with fewer photons (a direct correlation with the higher average noise in \textit{TESS}), and larger pixels compared to the point spread function (PSF).
This means centroids are noisier, and \textit{TESS} may not see a centroid shift on an object for which \textit{Kepler} was able to.
This discrepancy may also explain why we initially found that adding centroids caused problems with model training.
Another reason is the increased presence of NaNs in the input data arrays, due partly to the shorter baseline and decrease in the minimum number of transits to two compared to three in \textit{Kepler}.

Another source of the discrepancy is in noise in the labels. 
Although it may seem like simulated data would be easy to identify true signals vs no signal, this is not necessarily true. 
For example, single transits and eclipses were frequently detected by \textit{TESS}, with an incorrect period and/or with a second transit detected corresponding to some systematic noise or gap.
Our correlation metric would, in these cases, discard this as a "near miss". However, to the neural network, it is indeed seeing the signal of an astrophysical source.

A manual inspection of those candidates predicted to be planets by the 3-class model reveals that 69\% of those 284 objects with "unknown" ground truth were in fact co-incident with planetary injections.
Of those, 44\% came from monotransits, 25\% came from period confusion in multi-planet systems (e.g. a period near resonant with two or more planets, producing a planet-like phase-folded transit in combination), and the other 31\% from other reasons such as single or half-transits left by the first iteration of transit detection which are then identified at the wrong period in the second search; and transits close to, but not at, the correct period which become "smeared" in phase-space.
Immediately including these in the correctly identified boxes improves planet accuracy in the cross-validation results for the 3-class model from 90.3 to 95.7\% and the average precision across all classes to 95.0\%, nearly matching that of \citet{ansdell2018}. A similar increase would be expected in the ensemble test data.
However this poses a question as to exactly what constitutes a bona fide planet detection, and whether planets on missed periods constitute a true detection or not.
One improvement might be to apply a continuous label, from the degree of correlation between injection and recovered signal, rather than a pure binary label, however this is beyond the scope of this work.

\subsection{Comments on Multiclass and Binary models}
We attempted to train both binary and multi-class models partly because we assumed that the simplicity of a binary model may improve performance. 
We thought that a multi-class model, with specific knowledge of the source of the possible false positive contaminant, may aid planet follow-up. 
For example, class confusion between planet and background eclipsing binary may lead to the need for high-resolution imagery, whereas confusion between a planet and a non-astrophysical signal may lead to follow-up photometric observations of a future transit.

However, our results suggest that there is minimal difference between binary and multi-class models, as Figures \ref{fig:compPR} suggests.
In fact, the highest average precision on planets in the binary and 3-class models were equal at A.P.$=95.6$\%.
And this is likely to be even higher if monotransits and other near-miss planetary signals are included as true positives.

\begin{figure}
\begin{centering}
\includegraphics[width=\columnwidth]{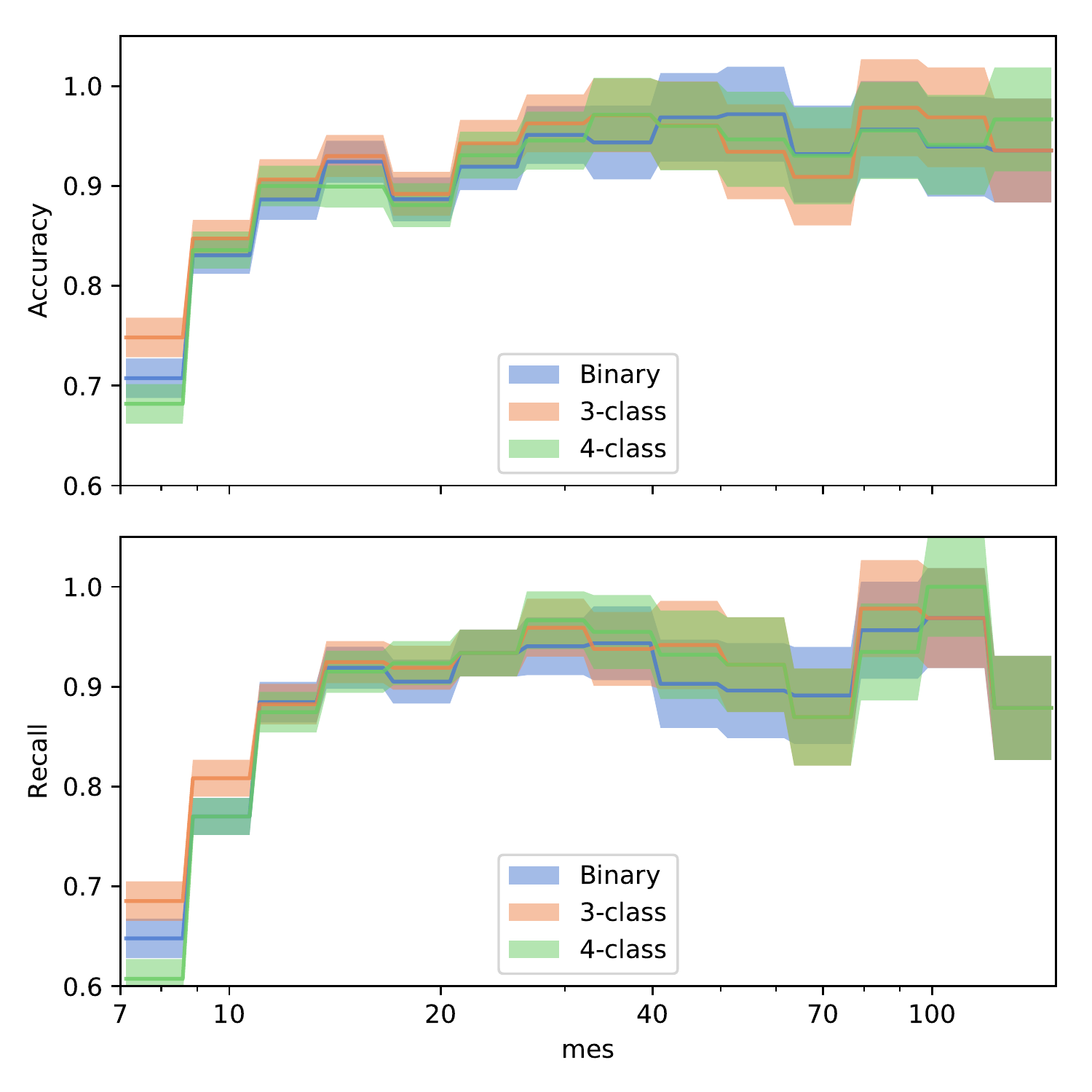}
\caption{Comparison of recall and accuracy for planets as a function of SNR across all three models. Note the y-scale has been rescaled between 60 and 100\%.}
\label{fig:compMES}
\end{centering}
\end{figure}

\begin{figure}
\begin{centering}
\includegraphics[width=\columnwidth]{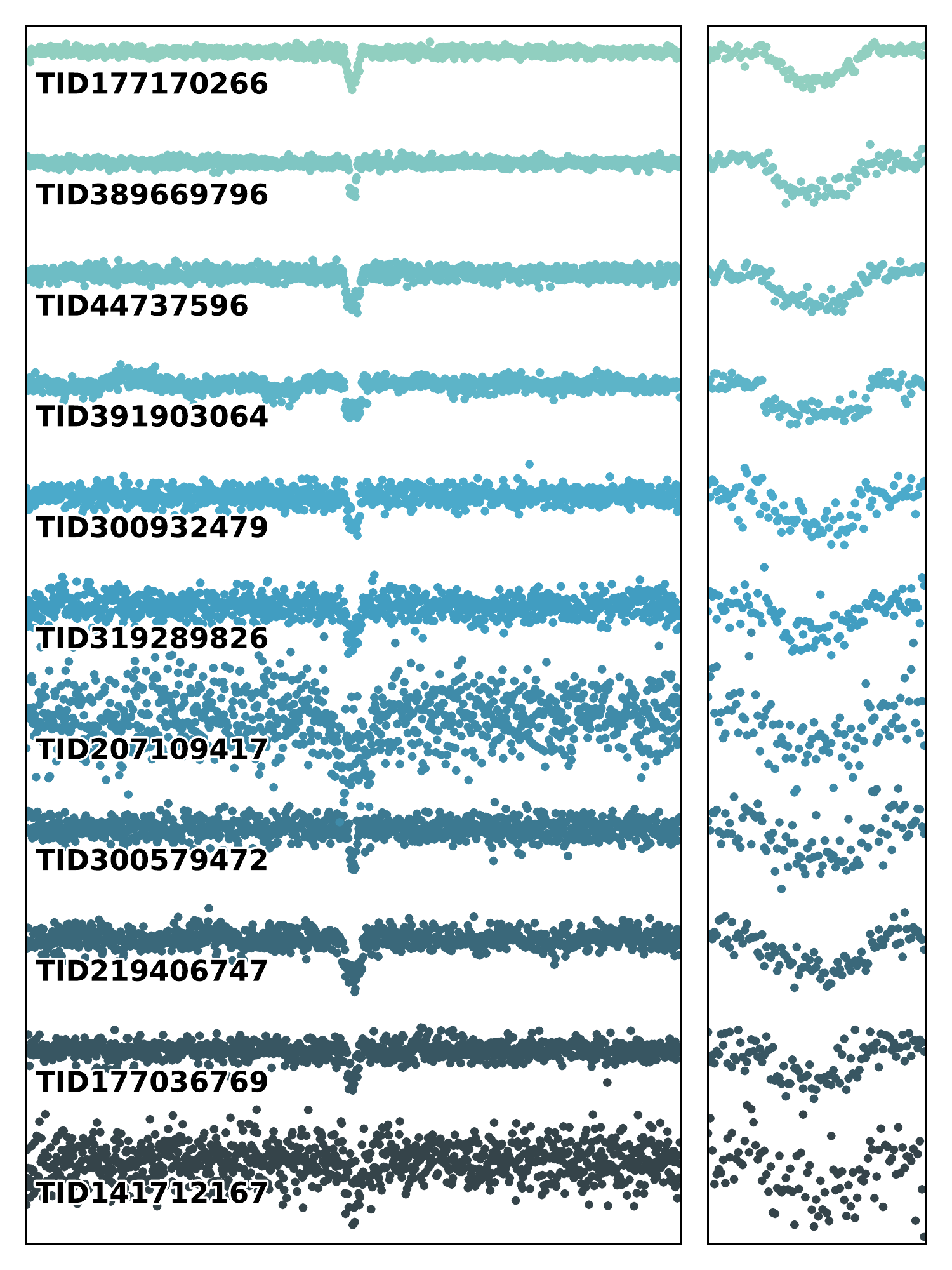}
\caption{A small selection of TCEs that have not been classified as TOIs but nonetheless appear to be good planet candidates. These come from a manual search for planet-like signals amongst those 200 TCEs with high $p_{\rm pl}$ predictions from our CNN models. The left panel shows "global" view for the whole phase while the right panel shows the "local" view. These are sorted by transit SNR. See Table A.\ref{tab:realpreds} for full information on all predicted planets.}
\label{fig:NewPLs}
\end{centering}
\end{figure}

Figure \ref{fig:compMES} also shows how all three models perform worse at classifying planets at lower SNR, with only between 60 and 70\% of planets with $7 < {\rm SNR} <8.5$ detected.
This decrease is expected, as the threshold for detections which become TCEs ($7.2\sigma$) is set such that the fraction of signals at this threshold which are from real astrophysical sources (both planets, EBs and BEBs) is 50\%.
Hence far fewer than 50\% of TCEs with MES$\sim 7.1\sigma$ are likely to be planets, hence our value of up to $\sim70\%$ shows a marked improvement.

Where multiclass models did have a noticeable negative effect was on the accuracy of identifying EB and BEB objects in the 4-class model. 
When normalised by depth and duration, blended binaries have identical shapes to eclipsing binaries, although often at lower signal to noise.
Hence these two classes were frequently confused, as Figure \ref{fig:4classCM} shows, leading to lower average precision for the model as a whole.

Another noticeable result from the multiclass models is that our model performs best at identifying non-astrophysical false positives.
This is especially true at low signal to noise (see Figure \ref{fig:MESplot}) where the recall and accuracy on these unknown signals actually increase.
While this may suggest that the model is, in some ways, gaming the system and applying the class of "unknown" for all transit events with high noise, where such signals dominate.
However, the accuracy of the classification also suggests that the model is learning the systematic noise inherent to the data, and is therefore able to separate these signals from the astrophysical classes.
This in itself is extremely important as often these systematic noise sources are varied and un-modellable, and therefore difficult to distinguish using classical techniques.

\subsection{Real TESS data}
Directly applying a model trained on simulated data to real data is a risky strategy. 
Although the planet and EB signals are likely to look physically similar, the characteristics of the systematic noise is likely extremely different.
However, a recall of 61\% on the TOI list is a relatively good sign that the model is transferable.
Especially given the different techniques and even pipeline used to create the TOI list, and given the as-yet unknown ground truth of those targets in the TOI list.
Indeed, our models suggest a handful of TOIs are indeed most likely to be astrophysical false positives.

Our model also predicted nearly 100 further TCEs as having high ($p_{\rm pl}>0.95$) planetary class probabilities.
These include a 760ppm signal from HD 55820 (TIC391903064), a 4ppt transit on HD 270135 (TIC389669796) and a 500ppm signal from TIC207109417.
Such targets are ripe for follow-up, and we hope future vetting by improved models will confirm these signals and identify even more.
Unfortunately, a quick look at those predicted planets also reveals many clear binaries, although the majority have transit-like eclipse shapes due to a large radius ratio, such as binary companions of giant stars.
This may be due to our input training set lacking objects of this nature.

Clearly our recall and accuracy on planet candidates in the simulated data is not matched when applying that to real data. 
However, without full knowledge of the ground truth in the real \textit{TESS} data, assessing model performance will be intrinsically more challenging.
In order to best represent the realistic noise sources, one could inject and recover realistic transit signals in real \textit{TESS} data.
However, this work was started before real data was available, and performing injections and recovery in real \textit{TESS} light curves is beyond the scope of this paper. 
We intend to perform such a task in a future publication.

\section{Conclusion}
The classification of candidates in exoplanet transit surveys can be a long and labour-intensive process when manual techniques are used.
Neural network-based classifiers \texttt{Astronet} \citep{Shallue2018} and \texttt{exonet} \citep{ansdell2018} have proved themselves extremely accurate \citep[98\% average precision in ][]{ansdell2018} and, once trained, can classify potential planets extremely rapidly.
We set out to apply such models to \textit{TESS}-like data.

To do this, we followed the \texttt{Astronet} technique of using local \& global "views" of the phase-folded, binned photometric data for each candidate, as well as the improvements of \texttt{exonet} --- namely including the centroids and stellar parameters.
In order to improve results, we also added data augmentation by mimicking additional noise sources, and use balanced batch sampling to normalise the unequal number of samples of each class in training data.

Using 4 sectors of pixel-level simulations with injected planets and false positive populations (known as TSOP-301), we trained three models with varying numbers of source classes using cross validation.
We achieve average precision as high as 97.3\% with accuracy on planet populations as high as 91.8\%.
This is despite limitations when compared to those results using \textit{Kepler} data, such as lower-significance centroids, a large population of non-astrophysical false positives (which were partly removed in the \textit{Kepler} ML dataset), and a higher degree of confusion between real planet signals and noise due to the lowered threshold on the number of transits from three to two.
Indeed, when positives from confusion between planet injections, or monotransits identified at the wrong period are included in the "planet" class, accuracy rises to as high as 95.7\%.

We also show that our models perform well even at low-SNR with accuracy on planets as high as 75\% for signals with ${\rm SNR}<8.5$. 
Our use of multi-class models may also aid targeted follow-up observations to the most likely source of confusion.
The high accuracy in non-astrophysical false positives also suggest that our neural network is able to learn patterns in low-significance systematic noise sources. This could therefore push planet detection closer to the theoretical SNR limit than is possible with classical vetting techniques

Once these models were trained on simulated data, we applied them to real \textit{TESS} candidates from sectors 1 to 4. 
Although no ground truth exists to test the performance of this model, we recover more than 60\% of the currently identified TOI list, including more than 95\% of all planets identified before the mission.
We also identify 14 TOIs as likely false-positives.
However, the use of confirmed \textit{TESS} planets as a training set, plus injections of simulated transits into real flight data, would improve our confidence in such classifications. This will form the next step in this ongoing project.

\begin{acknowledgements}
Software used includes: Astropy \citep{AstroPy2013,AstroPy2018}, PyTorch \citep{Paszke2017}, Astronet \citep{Shallue2018}, Jupyter \citep{Kluyver:2016aa}, SciPy \citep{SciPy2001}, Scikit Learn \citep{scikit-learn}, Matplotlib \citep{Hunter2007}.

This work, along with \citet{ansdell2018}, are the result of the 2018 NASA Frontier Development Lab\footnote{\url{https://frontierdevelopmentlab.org/}} (FDL) exoplanet challenge, and we are extremely grateful to the organisors, mentors and sponsors for providing this opportunity.

This material is based upon work supported by Google Cloud.
MA also gratefully acknowledges the NVIDIA Corporation for donating the Quadro P6000 GPU used for this research. We thank Adam Lesnikowski, Noa Kel, Hamed Valizadegan and Yarin Gal for their helpful discussions. This paper includes data from the {\it TESS} mission; funding for the {\it TESS} mission is provided by the NASA Explorer Program. The data used in this paper were obtained from the Mikulski Archive for Space Telescopes (MAST). STScI is operated by the Association of Universities for Research in Astronomy, Inc., under NASA contract NAS5-26555. MA also acknowledges support from NSF grant AST-1518332 and NASA grants NNX15AC89G and NNX15AD95G/NEXSS. We acknowledge the use of public TESS Alert data from pipelines at the TESS Science Office and at the TESS Science Processing Operations Center. HPO acknowledges support from Centre National d'Etudes Spatiales (CNES) grant 131425-PLATO.
\end{acknowledgements}

\bibliographystyle{aa}
\bibliography{aanda.bib}

\begin{thebibliography}{41}
\expandafter\ifx\csname natexlab\endcsname\relax\def\natexlab#1{#1}\fi

\bibitem[{Ansdell {et~al.}(2018)Ansdell, Ioannou, Osborn, Sasdelli, Smith,
  Caldwell, Jenkins, RÃ€issi, Angerhausen, \& and}]{ansdell2018}
Ansdell, M., Ioannou, Y., Osborn, H.~P., {et~al.} 2018, The Astrophysical
  Journal, 869, L7

\bibitem[{Armstrong {et~al.}(2018)Armstrong, G{\"u}nther, McCormac, Smith,
  Bayliss, Bouchy, Burleigh, Casewell, Eigm{\"u}ller, Gillen,
  {et~al.}}]{armstrong2018automatic}
Armstrong, D.~J., G{\"u}nther, M.~N., McCormac, J., {et~al.} 2018, Monthly
  Notices of the Royal Astronomical Society

\bibitem[{{Astropy Collaboration} {et~al.}(2018){Astropy Collaboration},
  {Price-Whelan}, {Sip{\H o}cz}, {G{\"u}nther}, {Lim}, {Crawford}, {Conseil},
  {Shupe}, {Craig}, {Dencheva}, {Ginsburg}, {VanderPlas}, {Bradley},
  {P{\'e}rez-Su{\'a}rez}, {de Val-Borro}, {Aldcroft}, {Cruz}, {Robitaille},
  {Tollerud}, {Ardelean}, {Babej}, {Bach}, {Bachetti}, {Bakanov}, {Bamford},
  {Barentsen}, {Barmby}, {Baumbach}, {Berry}, {Biscani}, {Boquien}, {Bostroem},
  {Bouma}, {Brammer}, {Bray}, {Breytenbach}, {Buddelmeijer}, {Burke},
  {Calderone}, {Cano Rodr{\'{\i}}guez}, {Cara}, {Cardoso}, {Cheedella},
  {Copin}, {Corrales}, {Crichton}, {D'Avella}, {Deil}, {Depagne}, {Dietrich},
  {Donath}, {Droettboom}, {Earl}, {Erben}, {Fabbro}, {Ferreira}, {Finethy},
  {Fox}, {Garrison}, {Gibbons}, {Goldstein}, {Gommers}, {Greco}, {Greenfield},
  {Groener}, {Grollier}, {Hagen}, {Hirst}, {Homeier}, {Horton}, {Hosseinzadeh},
  {Hu}, {Hunkeler}, {Ivezi{\'c}}, {Jain}, {Jenness}, {Kanarek}, {Kendrew},
  {Kern}, {Kerzendorf}, {Khvalko}, {King}, {Kirkby}, {Kulkarni}, {Kumar},
  {Lee}, {Lenz}, {Littlefair}, {Ma}, {Macleod}, {Mastropietro}, {McCully},
  {Montagnac}, {Morris}, {Mueller}, {Mumford}, {Muna}, {Murphy}, {Nelson},
  {Nguyen}, {Ninan}, {N{\"o}the}, {Ogaz}, {Oh}, {Parejko}, {Parley}, {Pascual},
  {Patil}, {Patil}, {Plunkett}, {Prochaska}, {Rastogi}, {Reddy Janga},
  {Sabater}, {Sakurikar}, {Seifert}, {Sherbert}, {Sherwood-Taylor}, {Shih},
  {Sick}, {Silbiger}, {Singanamalla}, {Singer}, {Sladen}, {Sooley},
  {Sornarajah}, {Streicher}, {Teuben}, {Thomas}, {Tremblay}, {Turner},
  {Terr{\'o}n}, {van Kerkwijk}, {de la Vega}, {Watkins}, {Weaver}, {Whitmore},
  {Woillez}, {Zabalza}, \& {Astropy Contributors}}]{AstroPy2018}
{Astropy Collaboration}, {Price-Whelan}, A.~M., {Sip{\H o}cz}, B.~M., {et~al.}
  2018, \aj, 156, 123

\bibitem[{{Astropy Collaboration} {et~al.}(2013){Astropy Collaboration},
  {Robitaille}, {Tollerud}, {Greenfield}, {Droettboom}, {Bray}, {Aldcroft},
  {Davis}, {Ginsburg}, {Price-Whelan}, {Kerzendorf}, {Conley}, {Crighton},
  {Barbary}, {Muna}, {Ferguson}, {Grollier}, {Parikh}, {Nair}, {Unther},
  {Deil}, {Woillez}, {Conseil}, {Kramer}, {Turner}, {Singer}, {Fox}, {Weaver},
  {Zabalza}, {Edwards}, {Azalee Bostroem}, {Burke}, {Casey}, {Crawford},
  {Dencheva}, {Ely}, {Jenness}, {Labrie}, {Lim}, {Pierfederici}, {Pontzen},
  {Ptak}, {Refsdal}, {Servillat}, \& {Streicher}}]{AstroPy2013}
{Astropy Collaboration}, {Robitaille}, T.~P., {Tollerud}, E.~J., {et~al.} 2013,
  \aap, 558, A33

\bibitem[{Barclay {et~al.}(2018)Barclay, Pepper, \&
  Quintana}]{barclay2018revised}
Barclay, T., Pepper, J., \& Quintana, E.~V. 2018, arXiv preprint
  arXiv:1804.05050

\bibitem[{Batalha {et~al.}(2013)Batalha, Rowe, Bryson, Barclay, Burke,
  Caldwell, Christiansen, Mullally, Thompson, Brown,
  {et~al.}}]{batalha2013planetary}
Batalha, N.~M., Rowe, J.~F., Bryson, S.~T., {et~al.} 2013, The Astrophysical
  Journal Supplement Series, 204, 24

\bibitem[{{Burke} {et~al.}(2014){Burke}, {Bryson}, {Mullally}, {Rowe},
  {Christiansen}, {Thompson}, {Coughlin}, {Haas}, {Batalha}, {Caldwell},
  {Jenkins}, {Still}, {Barclay}, {Borucki}, {Chaplin}, {Ciardi}, {Clarke},
  {Cochran}, {Demory}, {Esquerdo}, {Gautier}, {Gilliland }, {Girouard},
  {Havel}, {Henze}, {Howell}, {Huber}, {Latham}, {Li}, {Morehead}, {Morton},
  {Pepper}, {Quintana}, {Ragozzine}, {Seader}, {Shah}, {Shporer}, {Tenenbaum},
  {Twicken}, \& {Wolfgang}}]{2014ApJS..210...19B}
{Burke}, C.~J., {Bryson}, S.~T., {Mullally}, F., {et~al.} 2014, The
  Astrophysical Journal Supplement Series, 210, 19

\bibitem[{Chawla {et~al.}(2004)Chawla, Japkowicz, \& Kotcz}]{chawla2004special}
Chawla, N.~V., Japkowicz, N., \& Kotcz, A. 2004, ACM Sigkdd Explorations
  Newsletter, 6, 1

\bibitem[{Crossfield {et~al.}(2018)Crossfield, Guerrero, David, Quinn,
  Feinstein, Huang, Yu, Collins, Fulton, Benneke,
  {et~al.}}]{crossfield2018tess}
Crossfield, I.~J., Guerrero, N., David, T., {et~al.} 2018, arXiv preprint
  arXiv:1806.03127

\bibitem[{de~Val-Borro {et~al.}(2016)de~Val-Borro, Bakos, Brahm, Hartman,
  Espinoza, Penev, Ciceri, Jord{\'a}n, Bhatti, Csubry, {et~al.}}]{de2016hats}
de~Val-Borro, M., Bakos, G., Brahm, R., {et~al.} 2016, The Astronomical
  Journal, 152, 161

\bibitem[{Dietterich(2000)}]{dietterich2000ensemble}
Dietterich, T.~G. 2000, in International workshop on multiple classifier
  systems, Springer, 1--15

\bibitem[{{Fausnaugh} {et~al.}(2018){Fausnaugh}, {Huang}, {Glidden},
  {Guerrero}, \& {TESS Science Office}}]{2018AAS...23143909F}
{Fausnaugh}, M., {Huang}, X., {Glidden}, A., {Guerrero}, N., \& {TESS Science
  Office}. 2018, in American Astronomical Society Meeting Abstracts, Vol. 231,
  American Astronomical Society Meeting Abstracts \#231, 439.09

\bibitem[{Fischer {et~al.}(2012)Fischer, Schwamb, Schawinski, Lintott, Brewer,
  Giguere, Lynn, Parrish, Sartori, Simpson, {et~al.}}]{fischer2012planet}
Fischer, D.~A., Schwamb, M.~E., Schawinski, K., {et~al.} 2012, Monthly Notices
  of the Royal Astronomical Society, 419, 2900

\bibitem[{He \& Garcia(2008)}]{he2008learning}
He, H. \& Garcia, E.~A. 2008, IEEE Transactions on Knowledge \& Data
  Engineering, 1263

\bibitem[{Huang {et~al.}(2018{\natexlab{a}})Huang, Burt, Vanderburg,
  G{\"u}nther, Shporer, Dittmann, Winn, Wittenmyer, Sha, Kane,
  {et~al.}}]{huang2018tess}
Huang, C.~X., Burt, J., Vanderburg, A., {et~al.} 2018{\natexlab{a}}, arXiv
  preprint arXiv:1809.05967

\bibitem[{Huang {et~al.}(2018{\natexlab{b}})Huang, Shporer, Dragomir,
  Fausnaugh, Levine, Morgan, Nguyen, Ricker, Wall, Woods,
  {et~al.}}]{huang2018expected}
Huang, C.~X., Shporer, A., Dragomir, D., {et~al.} 2018{\natexlab{b}}, arXiv
  preprint arXiv:1807.11129

\bibitem[{Hunter(2007)}]{Hunter2007}
Hunter, J.~D. 2007, Computing In Science \& Engineering, 9, 90

\bibitem[{{Jenkins} {et~al.}(2002){Jenkins}, {Caldwell}, \&
  {Borucki}}]{Jenkins2002}
{Jenkins}, J.~M., {Caldwell}, D.~A., \& {Borucki}, W.~J. 2002, \apj, 564, 495

\bibitem[{{Jenkins} {et~al.}(2010){Jenkins}, {Chandrasekaran}, {McCauliff},
  {Caldwell}, {Tenenbaum}, {Li}, {Klaus}, {Cote}, \&
  {Middour}}]{2010SPIE.7740E..0DJ}
{Jenkins}, J.~M., {Chandrasekaran}, H., {McCauliff}, S.~D., {et~al.} 2010, in
  Society of Photo-Optical Instrumentation Engineers (SPIE) Conference Series,
  Vol. 7740, Software and Cyberinfrastructure for Astronomy, 77400D

\bibitem[{Jenkins {et~al.}(2018)Jenkins, Tenenbaum, Caldwell, Davies, Li,
  Morris, Rose, Smith, Ting, Twicken, {et~al.}}]{jenkins2018simulated}
Jenkins, J.~M., Tenenbaum, P., Caldwell, D.~A., {et~al.} 2018, Research Notes
  of the AAS, 2, 47

\bibitem[{{Jenkins} {et~al.}(2016){Jenkins}, {Twicken}, {McCauliff},
  {Campbell}, {Sanderfer}, {Lung}, {Mansouri-Samani}, {Girouard}, {Tenenbaum},
  {Klaus}, {Smith}, {Caldwell}, {Chacon}, {Henze}, {Heiges}, {Latham},
  {Morgan}, {Swade}, {Rinehart}, \& {Vanderspek}}]{2016SPIE.9913E..3EJ}
{Jenkins}, J.~M., {Twicken}, J.~D., {McCauliff}, S., {et~al.} 2016, in
  \procspie, Vol. 9913, Software and Cyberinfrastructure for Astronomy IV,
  99133E

\bibitem[{Jones {et~al.}(2001--)Jones, Oliphant, Peterson,
  {et~al.}}]{SciPy2001}
Jones, E., Oliphant, T., Peterson, P., {et~al.} 2001--, {SciPy}: Open source
  scientific tools for {Python}, [Online; accessed <today>]

\bibitem[{Kingma \& Ba(2014)}]{kingma2014adam}
Kingma, D.~P. \& Ba, J. 2014, arXiv preprint arXiv:1412.6980

\bibitem[{Kluyver {et~al.}(2016)Kluyver, Ragan-Kelley, P{\'e}rez, Granger,
  Bussonnier, Frederic, Kelley, Hamrick, Grout, Corlay, Ivanov, Avila, Abdalla,
  \& Willing}]{Kluyver:2016aa}
Kluyver, T., Ragan-Kelley, B., P{\'e}rez, F., {et~al.} 2016, in Positioning and
  Power in Academic Publishing: Players, Agents and Agendas, ed. F.~Loizides \&
  B.~Schmidt, IOS Press, 87 -- 90

\bibitem[{{Li} {et~al.}(2019){Li}, {Tenenbaum}, {Twicken}, {Burke}, {Jenkins},
  {Quintana}, {Rowe}, \& {Seader}}]{2019PASP..131b4506L}
{Li}, J., {Tenenbaum}, P., {Twicken}, J.~D., {et~al.} 2019, Publications of the
  Astronomical Society of the Pacific, 131, 024506

\bibitem[{McCauliff {et~al.}(2015)McCauliff, Jenkins, Catanzarite, Burke,
  Coughlin, Twicken, Tenenbaum, Seader, Li, \& Cote}]{mccauliff2015automatic}
McCauliff, S.~D., Jenkins, J.~M., Catanzarite, J., {et~al.} 2015, The
  Astrophysical Journal, 806, 6

\bibitem[{Mullally {et~al.}(2016)Mullally, Coughlin, Thompson, Christiansen,
  Burke, Clarke, \& Haas}]{mullally2016identifying}
Mullally, F., Coughlin, J.~L., Thompson, S.~E., {et~al.} 2016, Publications of
  the Astronomical Society of the Pacific, 128, 074502

\bibitem[{{Mullally} {et~al.}(2015){Mullally}, {Coughlin}, {Thompson}, {Rowe},
  {Burke}, {Latham}, {Batalha}, {Bryson}, {Christiansen}, {Henze}, {Ofir},
  {Quarles}, {Shporer}, {Van Eylen}, {Van Laerhoven}, {Shah}, {Wolfgang},
  {Chaplin}, {Xie}, {Akeson}, {Argabright}, {Bachtell}, {Barclay}, {Borucki},
  {Caldwell}, {Campbell}, {Catanzarite}, {Cochran}, {Duren}, {Fleming},
  {Fraquelli}, {Girouard}, {Haas}, {He{\l}miniak}, {Howell}, {Huber}, {Larson},
  {Gautier}, {Jenkins}, {Li}, {Lissauer}, {McArthur}, {Miller}, {Morris},
  {Patil-Sabale}, {Plavchan}, {Putnam}, {Quintana}, {Ramirez}, {Silva Aguirre},
  {Seader}, {Smith}, {Steffen}, {Stewart}, {Stober}, {Still}, {Tenenbaum},
  {Troeltzsch}, {Twicken}, \& {Zamudio}}]{2015ApJS..217...31M}
{Mullally}, F., {Coughlin}, J.~L., {Thompson}, S.~E., {et~al.} 2015, The
  Astrophysical Journal Supplement Series, 217, 31

\bibitem[{Paszke {et~al.}(2017)Paszke, Gross, Chintala, Chanan, Yang, DeVito,
  Lin, Desmaison, Antiga, \& Lerer}]{Paszke2017}
Paszke, A., Gross, S., Chintala, S., {et~al.} 2017, in NIPS-W

\bibitem[{Pedregosa {et~al.}(2011)Pedregosa, Varoquaux, Gramfort, Michel,
  Thirion, Grisel, Blondel, Prettenhofer, Weiss, Dubourg, Vanderplas, Passos,
  Cournapeau, Brucher, Perrot, \& Duchesnay}]{scikit-learn}
Pedregosa, F., Varoquaux, G., Gramfort, A., {et~al.} 2011, Journal of Machine
  Learning Research, 12, 2825

\bibitem[{{Ricker} {et~al.}(2014){Ricker}, {Winn}, {Vanderspek}, {Latham},
  {Bakos}, {Bean}, {Berta-Thompson}, {Brown}, {Buchhave}, {Butler}, {Butler},
  {Chaplin}, {Charbonneau}, {Christensen-Dalsgaard}, {Clampin}, {Deming},
  {Doty}, {De Lee}, {Dressing}, {Dunham}, {Endl}, {Fressin}, {Ge}, {Henning},
  {Holman}, {Howard}, {Ida}, {Jenkins}, {Jernigan}, {Johnson}, {Kaltenegger},
  {Kawai}, {Kjeldsen}, {Laughlin}, {Levine}, {Lin}, {Lissauer}, {MacQueen},
  {Marcy}, {McCullough}, {Morton}, {Narita}, {Paegert}, {Palle}, {Pepe},
  {Pepper}, {Quirrenbach}, {Rinehart}, {Sasselov}, {Sato}, {Seager},
  {Sozzetti}, {Stassun}, {Sullivan}, {Szentgyorgyi}, {Torres}, {Udry}, \&
  {Villasenor}}]{Ricker2014}
{Ricker}, G.~R., {Winn}, J.~N., {Vanderspek}, R., {et~al.} 2014, in \procspie,
  Vol. 9143, Space Telescopes and Instrumentation 2014: Optical, Infrared, and
  Millimeter Wave, 914320

\bibitem[{{Rowe} {et~al.}(2015){Rowe}, {Coughlin}, {Antoci}, {Barclay},
  {Batalha}, {Borucki}, {Burke}, {Bryson}, {Caldwell}, {Campbell},
  {Catanzarite}, {Christiansen}, {Cochran}, {Gilliland}, {Girouard}, {Haas},
  {He{\l}miniak}, {Henze}, {Hoffman}, {Howell}, {Huber}, {Hunter},
  {Jang-Condell}, {Jenkins}, {Klaus}, {Latham}, {Li}, {Lissauer}, {McCauliff},
  {Morris}, {Mullally}, {Ofir}, {Quarles}, {Quintana}, {Sabale}, {Seader},
  {Shporer}, {Smith}, {Steffen}, {Still}, {Tenenbaum}, {Thompson}, {Twicken},
  {Van Laerhoven}, {Wolfgang}, \& {Zamudio}}]{2015ApJS..217...16R}
{Rowe}, J.~F., {Coughlin}, J.~L., {Antoci}, V., {et~al.} 2015, The
  Astrophysical Journal Supplement Series, 217, 16

\bibitem[{Schanche {et~al.}(2018)Schanche, Cameron, H{\'e}brard, Nielsen,
  Triaud, Almenara, Alsubai, Anderson, Armstrong, Barros,
  {et~al.}}]{schanche2018machine}
Schanche, N., Cameron, A.~C., H{\'e}brard, G., {et~al.} 2018, Monthly Notices
  of the Royal Astronomical Society, 483, 5534

\bibitem[{{Seader} {et~al.}(2013){Seader}, {Tenenbaum}, {Jenkins}, \&
  {Burke}}]{2013ApJS..206...25S}
{Seader}, S., {Tenenbaum}, P., {Jenkins}, J.~M., \& {Burke}, C.~J. 2013, The
  Astrophysical Journal Supplement Series, 206, 25

\bibitem[{{Shallue} \& {Vanderburg}(2018)}]{Shallue2018}
{Shallue}, C.~J. \& {Vanderburg}, A. 2018, \aj, 155, 94

\bibitem[{Shporer {et~al.}(2018)Shporer, Wong, Huang, Line, Stassun, Fetherolf,
  Kane, Ricker, Latham, Seager, {et~al.}}]{shporer2018tess}
Shporer, A., Wong, I., Huang, C.~X., {et~al.} 2018, arXiv preprint
  arXiv:1811.06020

\bibitem[{Sullivan {et~al.}(2015)Sullivan, Winn, Berta-Thompson, Charbonneau,
  Deming, Dressing, Latham, Levine, McCullough, Morton,
  {et~al.}}]{sullivan2015transiting}
Sullivan, P.~W., Winn, J.~N., Berta-Thompson, Z.~K., {et~al.} 2015, The
  Astrophysical Journal, 809, 77

\bibitem[{{Twicken} {et~al.}(2018){Twicken}, {Catanzarite}, {Clarke},
  {Girouard}, {Jenkins}, {Klaus}, {Li}, {McCauliff}, {Seader}, {Tenenbaum},
  {Wohler}, {Bryson}, {Burke}, {Caldwell}, {Haas}, {Henze}, \&
  {Sanderfer}}]{2018PASP..130f4502T}
{Twicken}, J.~D., {Catanzarite}, J.~H., {Clarke}, B.~D., {et~al.} 2018,
  Publications of the Astronomical Society of the Pacific, 130, 064502

\bibitem[{Vanderspek {et~al.}(2018)Vanderspek, Huang, Vanderburg, Ricker,
  Latham, Seager, Winn, Jenkins, Burt, Dittmann, {et~al.}}]{vanderspek2018tess}
Vanderspek, R., Huang, C.~X., Vanderburg, A., {et~al.} 2018, arXiv preprint
  arXiv:1809.07242

\bibitem[{Wang {et~al.}(2018)}]{wang2018transiting}
Wang, S. {et~al.} 2018, arXiv preprint arXiv:1810.02341

\bibitem[{{Zucker} \& {Giryes}(2018)}]{Zucker2018}
{Zucker}, S. \& {Giryes}, R. 2018, \aj, 155, 147

\end{thebibliography}

\begin{appendix}

\numberwithin{table}{section}

\onecolumn
\begin{table}
    \caption{TCEs that correspond to detected \textit{TESS} Objects of Interest (TOIs), with the class predictions as given by our CNN models. These are ranked by the average planet score.}
    \label{tab:toipreds}
    \centering
    \tiny
        \begin{tabular}{|l|l|c|c|c|c|c|c|c|c|c|c|c|c|c|c|}
            \hline
            \multicolumn{7}{|c|}{\textbf{TCE info}} & \textbf{Binary} & \multicolumn{3}{|c|}{\textbf{3-class prediction}} & \multicolumn{4}{|c|}{\textbf{4-class prediction}} & All \\
            \hline
            TID & TOI & Sect & Period (d) & Epoch & $t_D$ (hr) & $\delta$ (ppm) & PL & UNK & PL & EB & UNK & PL & EB & BEB & PL \\
           \hline
  77031414 &  241.01 &        2 &    1.387 &  1355.196 &  2.290 &   14260 &      0.998 &          0.000 &         1.000 &         0.000 &          0.000 &         1.000 &         0.000 &            0.0 &      0.999 \\
 122612091 &  264.01 &      3 4 &    2.217 &  1387.831 &  3.846 &    4180 &      0.997 &          0.000 &         1.000 &         0.000 &          0.000 &         1.000 &         0.000 &            0.0 &      0.999 \\
   1129033 &  398.01 &        4 &    1.360 &  1410.985 &  2.156 &   16380 &      0.997 &          0.000 &         1.000 &         0.000 &          0.000 &         1.000 &         0.000 &            0.0 &      0.999 \\
 290131778 &  123.01 &        1 &    3.309 &  1325.375 &  5.672 &    3230 &      0.997 &          0.000 &         1.000 &         0.000 &          0.000 &         1.000 &         0.000 &            0.0 &      0.999 \\
 144065872 &  105.01 &        1 &    2.185 &  1326.506 &  2.865 &   11840 &      0.997 &          0.000 &         1.000 &         0.000 &          0.000 &         1.000 &         0.000 &            0.0 &      0.999 \\
 230982885 &  195.01 &        2 &    2.073 &  1355.490 &  2.588 &   13510 &      0.997 &          0.000 &         1.000 &         0.000 &          0.000 &         1.000 &         0.000 &            0.0 &      0.999 \\
 184240683 &  250.01 &        2 &    1.628 &  1355.508 &  2.376 &   13800 &      0.997 &          0.000 &         1.000 &         0.000 &          0.000 &         1.000 &         0.000 &            0.0 &      0.999 \\
 267263253 &  135.01 &        1 &    4.127 &  1325.784 &  4.495 &   10070 &      0.997 &          0.000 &         1.000 &         0.000 &          0.000 &         1.000 &         0.000 &            0.0 &      0.999 \\
  25375553 &  143.01 &        1 &    2.311 &  1325.582 &  3.416 &    6950 &      0.996 &          0.000 &         1.000 &         0.000 &          0.000 &         1.000 &         0.000 &            0.0 &      0.999 \\
 388104525 &  112.01 &    1 2 3 &    2.500 &  1327.410 &  2.877 &   14940 &      0.996 &          0.000 &         1.000 &         0.000 &          0.000 &         1.000 &         0.000 &            0.0 &      0.999 \\
  25155310 &  114.01 &    1 2 3 &    3.289 &  1327.521 &  3.416 &    7180 &      0.996 &          0.000 &         1.000 &         0.000 &          0.000 &         1.000 &         0.000 &            0.0 &      0.999 \\
  38846515 &  106.01 &    1 2 3 &    2.849 &  1326.745 &  3.785 &    7500 &      0.996 &          0.000 &         1.000 &         0.000 &          0.000 &         1.000 &         0.000 &            0.0 &      0.999 \\
 422655579 &  388.01 &        4 &    2.903 &  1413.143 &  5.048 &    4660 &      0.995 &          0.000 &         1.000 &         0.000 &          0.000 &         1.000 &         0.000 &            0.0 &      0.998 \\
 402026209 &  232.01 &        2 &    1.338 &  1355.185 &  2.157 &   27440 &      0.995 &          0.000 &         1.000 &         0.000 &          0.000 &         1.000 &         0.000 &            0.0 &      0.998 \\
 231670397 &  104.01 &        1 &    4.087 &  1327.673 &  5.597 &    3610 &      0.993 &          0.000 &         1.000 &         0.000 &          0.000 &         1.000 &         0.000 &            0.0 &      0.998 \\
 149603524 &  102.01 &    1 2 3 &    4.412 &  1326.079 &  3.779 &   14030 &      0.992 &          0.000 &         1.000 &         0.000 &          0.000 &         1.000 &         0.000 &            0.0 &      0.997 \\
  92352620 &  107.01 &        1 &    3.950 &  1328.299 &  4.580 &   12930 &      0.990 &          0.000 &         1.000 &         0.000 &          0.000 &         1.000 &         0.000 &            0.0 &      0.997 \\
 336732616 &  103.01 &        1 &    3.548 &  1327.253 &  3.488 &   10400 &      0.989 &          0.000 &         1.000 &         0.000 &          0.000 &         1.000 &         0.000 &            0.0 &      0.996 \\
 166836920 &  267.01 &        3 &    5.752 &  1387.960 &  5.386 &    5390 &      0.986 &          0.000 &         1.000 &         0.000 &          0.000 &         1.000 &         0.000 &            0.0 &      0.995 \\
 281459670 &  110.01 &      1 2 &    3.174 &  1328.040 &  2.723 &   15600 &      0.984 &          0.000 &         1.000 &         0.000 &          0.000 &         1.000 &         0.000 &            0.0 &      0.995 \\
 170634116 &  413.01 &        4 &    3.662 &  1412.892 &  3.784 &   12460 &      0.980 &          0.000 &         1.000 &         0.000 &          0.000 &         1.000 &         0.000 &            0.0 &      0.993 \\
 257567854 &  403.01 &        4 &    3.533 &  1411.902 &  3.443 &   11280 &      0.980 &          0.000 &         1.000 &         0.000 &          0.000 &         1.000 &         0.000 &            0.0 &      0.993 \\
 238176110 &  116.01 &        1 &    2.799 &  1326.689 &  2.365 &   16850 &      0.978 &          0.000 &         1.000 &         0.000 &          0.000 &         1.000 &         0.000 &            0.0 &      0.993 \\
 260609205 &  219.01 &    1 2 3 &    4.462 &  1328.755 &  5.410 &   21140 &      0.976 &          0.000 &         1.000 &         0.000 &          0.000 &         1.000 &         0.000 &            0.0 &      0.992 \\
 183537452 &  192.01 &        2 &    3.923 &  1356.415 &  2.616 &   11500 &      0.968 &          0.000 &         1.000 &         0.000 &          0.000 &         1.000 &         0.000 &            0.0 &      0.989 \\
 403224672 &  141.01 &        1 &    1.008 &  1325.539 &  1.499 &     220 &      0.967 &          0.000 &         1.000 &         0.000 &          0.000 &         1.000 &         0.000 &            0.0 &      0.989 \\
 219253008 &  268.01 &      4 3 &    5.066 &  1415.524 &  6.158 &    1880 &      0.964 &          0.000 &         1.000 &         0.000 &          0.000 &         1.000 &         0.000 &            0.0 &      0.988 \\
  97409519 &  113.01 &        1 &    3.373 &  1327.053 &  2.640 &   17150 &      0.964 &          0.000 &         1.000 &         0.000 &          0.000 &         1.000 &         0.000 &            0.0 &      0.988 \\
 204376737 &  231.01 &        2 &    3.361 &  1357.395 &  2.579 &   24220 &      0.963 &          0.000 &         1.000 &         0.000 &          0.000 &         1.000 &         0.000 &            0.0 &      0.988 \\
 183979262 &  183.01 &  1 2 1 2 &    3.431 &  1326.104 &  3.130 &    8980 &      0.962 &          0.001 &         0.999 &         0.000 &          0.003 &         0.996 &         0.001 &            0.0 &      0.986 \\
  35857242 &  400.01 &        4 &    3.635 &  1413.315 &  4.005 &    8890 &      0.957 &          0.000 &         1.000 &         0.000 &          0.000 &         1.000 &         0.000 &            0.0 &      0.986 \\
 234994474 &  134.01 &        1 &    1.401 &  1326.033 &  1.261 &     560 &      0.956 &          0.000 &         0.999 &         0.001 &          0.000 &         1.000 &         0.000 &            0.0 &      0.985 \\
            \hline
            \multicolumn{16}{|c|}{ 201 rows }\\
            \hline
        \end{tabular}
\end{table}

\begin{table}
    \tiny
        \caption{TCEs which do not correspond to TOIs, but which our model gives a high average score for the planet class.}
        \label{tab:realpreds}
        \centering
        \begin{tabular}{|l|l|c|c|c|c|c|c|c|c|c|c|c|c|c|}
            \hline
            \multicolumn{6}{|c|}{\textbf{TCE info}} & \textbf{Binary} & \multicolumn{3}{|c|}{\textbf{3-class prediction}} & \multicolumn{4}{|c|}{\textbf{4-class prediction}} & All \\
            \hline
            TID & Sect & Period (d) & Epoch & $t_D$ (hr) & $\delta$ (ppm) &  PL & UNK & PL & EB & UNK & PL & EB & BEB & PL \\
            \hline
  92352621 &       1 &    3.950 &  1328.299 &  4.560 &   14230 &      0.997 &          0.000 &         1.000 &         0.000 &          0.000 &         1.000 &         0.000 &          0.000 &      0.999 \\
 139256217 &       1 &    2.199 &  1325.350 &  3.821 &   18410 &      0.997 &          0.000 &         1.000 &         0.000 &          0.000 &         1.000 &         0.000 &          0.000 &      0.999 \\
 259470701 &       4 &    2.543 &  1412.514 &  3.845 &   17670 &      0.995 &          0.000 &         1.000 &         0.000 &          0.000 &         1.000 &         0.000 &          0.000 &      0.998 \\
 231275247 &       2 &    4.008 &  1354.114 &  4.995 &   21100 &      0.995 &          0.000 &         1.000 &         0.000 &          0.000 &         1.000 &         0.000 &          0.000 &      0.998 \\
 272357134 &       1 &    4.197 &  1328.549 &  5.810 &   17080 &      0.994 &          0.000 &         1.000 &         0.000 &          0.000 &         1.000 &         0.000 &          0.000 &      0.998 \\
 272357134 &   1 2 3 &    4.197 &  1328.549 &  5.810 &   17030 &      0.992 &          0.000 &         1.000 &         0.000 &          0.000 &         1.000 &         0.000 &          0.000 &      0.997 \\
 272357134 &     1 2 &    4.197 &  1328.549 &  5.811 &   16990 &      0.992 &          0.000 &         1.000 &         0.000 &          0.000 &         1.000 &         0.000 &          0.000 &      0.997 \\
 237342298 &       3 &    2.692 &  1386.312 &  4.341 &   31710 &      0.991 &          0.000 &         1.000 &         0.000 &          0.000 &         1.000 &         0.000 &          0.000 &      0.997 \\
 231275247 &       3 &    4.008 &  1386.177 &  5.031 &   21460 &      0.991 &          0.000 &         1.000 &         0.000 &          0.000 &         1.000 &         0.000 &          0.000 &      0.997 \\
 272357134 &       4 &    4.197 &  1412.487 &  5.797 &   17060 &      0.991 &          0.000 &         1.000 &         0.000 &          0.000 &         1.000 &         0.000 &          0.000 &      0.997 \\
 272357134 &       3 &    4.197 &  1387.306 &  5.813 &   17090 &      0.990 &          0.000 &         1.000 &         0.000 &          0.000 &         1.000 &         0.000 &          0.000 &      0.997 \\
 272357134 &       2 &    4.197 &  1357.927 &  5.812 &   16860 &      0.989 &          0.000 &         1.000 &         0.000 &          0.000 &         1.000 &         0.000 &          0.000 &      0.996 \\
 231275247 &   1 2 3 &    4.008 &  1354.114 &  5.009 &   21170 &      0.988 &          0.000 &         1.000 &         0.000 &          0.000 &         1.000 &         0.000 &          0.000 &      0.996 \\
 349308842 &       4 &    3.582 &  1410.985 &  4.156 &   32620 &      0.985 &          0.000 &         1.000 &         0.000 &          0.000 &         1.000 &         0.000 &          0.000 &      0.995 \\
 170749770 &       4 &    5.306 &  1412.463 &  6.497 &   16800 &      0.985 &          0.000 &         1.000 &         0.000 &          0.000 &         1.000 &         0.000 &          0.000 &      0.995 \\
 159984211 &       3 &    3.660 &  1386.931 &  6.411 &   10150 &      0.982 &          0.000 &         1.000 &         0.000 &          0.000 &         1.000 &         0.000 &          0.000 &      0.994 \\
 237342298 &       4 &    2.692 &  1413.233 &  4.334 &   30650 &      0.982 &          0.000 &         1.000 &         0.000 &          0.000 &         1.000 &         0.000 &          0.000 &      0.994 \\
 159984211 &       4 &    3.660 &  1412.553 &  6.462 &   10990 &      0.980 &          0.000 &         1.000 &         0.000 &          0.000 &         1.000 &         0.000 &          0.000 &      0.993 \\
 350480660 &       4 &    4.464 &  1413.136 &  3.759 &   27390 &      0.980 &          0.000 &         1.000 &         0.000 &          0.000 &         1.000 &         0.000 &          0.000 &      0.993 \\
 350480660 &       3 &    4.464 &  1386.354 &  3.739 &   27110 &      0.976 &          0.000 &         1.000 &         0.000 &          0.000 &         1.000 &         0.000 &          0.000 &      0.992 \\
  64108432 &       4 &    2.779 &  1413.462 &  3.579 &   17940 &      0.970 &          0.000 &         1.000 &         0.000 &          0.000 &         1.000 &         0.000 &          0.000 &      0.990 \\
 115115136 &       1 &    3.962 &  1327.888 &  4.558 &   32680 &      0.968 &          0.000 &         1.000 &         0.000 &          0.000 &         1.000 &         0.000 &          0.000 &      0.989 \\
 268529943 &       3 &    4.301 &  1388.935 &  3.768 &   20290 &      0.967 &          0.000 &         1.000 &         0.000 &          0.000 &         1.000 &         0.000 &          0.000 &      0.989 \\
 167554898 &       4 &    4.453 &  1413.030 &  3.887 &    6540 &      0.965 &          0.000 &         1.000 &         0.000 &          0.000 &         1.000 &         0.000 &          0.000 &      0.988 \\
 260756218 &       1 &    1.955 &  1325.391 &  2.300 &   26760 &      0.941 &          0.000 &         1.000 &         0.000 &          0.000 &         1.000 &         0.000 &          0.000 &      0.980 \\
 422844353 &       4 &    6.635 &  1413.205 &  4.769 &   18370 &      0.920 &          0.000 &         1.000 &         0.000 &          0.000 &         1.000 &         0.000 &          0.000 &      0.973 \\
 270380593 &       4 &    3.836 &  1414.136 &  2.390 &    5580 &      0.911 &          0.000 &         1.000 &         0.000 &          0.000 &         1.000 &         0.000 &          0.000 &      0.970 \\
 391903064 &     1 2 &    4.666 &  1326.494 &  3.005 &     760 &      0.901 &          0.000 &         1.000 &         0.000 &          0.011 &         0.989 &         0.000 &          0.000 &      0.964 \\
 178242590 &       4 &    1.302 &  1411.139 &  2.727 &   15470 &      0.990 &          0.000 &         1.000 &         0.000 &          0.000 &         0.890 &         0.110 &          0.000 &      0.960 \\
 260304277 &     1 2 &    0.513 &  1325.773 &  1.100 &     180 &      0.863 &          0.000 &         1.000 &         0.000 &          0.000 &         0.999 &         0.000 &          0.001 &      0.954 \\
 234518605 &     1 2 &    5.679 &  1326.532 &  4.175 &   62410 &      0.859 &          0.000 &         1.000 &         0.000 &          0.000 &         0.998 &         0.002 &          0.000 &      0.953 \\
 164752991 &       3 &    1.222 &  1386.808 &  1.225 &    2930 &      0.918 &          0.004 &         0.969 &         0.027 &          0.000 &         0.970 &         0.000 &          0.030 &      0.952 \\
            \hline
            \multicolumn{15}{|c|}{ 210 rows }\\
            \hline
        \end{tabular}
\end{table}


\end{appendix}


\end{document}